\documentclass[a4paper,fleqn,usenatbib]{mnras}
\usepackage{hyperref}
\usepackage{subfig}
\usepackage{subfloat}
\usepackage{graphicx}
\usepackage{bm}
\usepackage{ae,aecompl}
\usepackage[T1]{fontenc}
\usepackage{aecompl}

\title[Search for a planet sculpting HD~115600 disc]{Numerical search for a potential planet sculpting the young disc of HD~115600}
\author[E. Thilliez et al.]{E. Thilliez$^{1}$\thanks{e-mail: ethilliez@astro.swin.edu.au}, S. T. Maddison$^1$\\
$^1$Centre for Astrophysics and Supercomputing, Swinburne University of Technology, Hawthorn, VIC 3122, Australia}

\date{Released 2016 Xxxxx XX}

\pubyear{2016}

\begin{document}
\label{firstpage}
\pagerange{\pageref{firstpage}--\pageref{lastpage}}
\maketitle

\begin{abstract}
Radial and azimuthal features (such as disc offsets and eccentric rings) seen in high resolution images of debris discs, provide us with the unique opportunity of finding potential planetary companions which betray their presence by gravitationally sculpting such asymmetric features. The young debris disc around HD~115600, imaged recently by the \emph{Gemini Planet Imager}, is such a disc with an eccentricity $e \sim 0.1-0.2$ and a projected offset from the star of $\sim 4$~AU. Using our modified N-body code which incorporates radiation forces, we firstly  aim to determine the orbit of a hidden planetary companion potentially responsible for shaping the disc. We run a suite of simulations covering a broad range of planetary parameters using a \emph{Monte Carlo Markov Chain} sampling method and create synthetic images from which we extract the geometric disc parameters to be compared with the observed and model-derived quantities. We then repeat the study using a traditional grid to explore the planetary parameter space and aim secondly to compare the efficiency of both sampling methods. We find a planet of 7.8 Jupiter mass orbiting at 30 AU with an eccentricity of $e=0.2$ to be the best fit to the observations of HD~115600. Technically, such planet has a contrast detectable by direct imaging, however the system's orientation does not favour such detection. In this study, at equal number of explored planetary configurations, the Monte Carlo Markov Chain not only converges faster but provides a better fit than a traditional grid.
\end{abstract}

\section{Introduction}
Planets can gravitationally perturb debris discs by various dynamical processes, such as secular interactions, where an eccentric or inclined planet can force the disc eccentricity or inclination \citep{1999ApJ...527..918W}, or resonance interactions, where the planet traps dust at a specific location, resulting in the creation of dust clumps in the disc \citep{2002ApJ...578L.149Q}. These processes inducing eccentricity, a disc position offset with respect to the star, or clumps into the disc can result in brightness asymmetries. Due to the limitations in the detection techniques, most of the confirmed exoplanets are located within 10~AU of their host star, and potential planets located beyond this limit (beyond Saturn in our solar system) remain undetected. However, even if those distant planets are too small to be detected with our current telescopes, they can still leave an observational signature by gravitationally perturbing the dust of their debris disc. Therefore investigating the dynamical relationship between debris discs and exoplanets can not only provide some insights on the origin of debris disc asymmetries, but also provides clues to the presence of hidden planets in the outer part of stellar systems, a region currently difficult to observe.

HD~115600 is a young F2 type star at a distance of 110~pc in the Scorpius--Centaurus Association \citep{2007A&A...474..653V} of age $\sim$15 Myr \citep{2012ApJ...746..154P}. Although an IR excess was detected by \cite{2011ApJ...738..122C} using \textit{Spitzer/MIPS}, its debris disc was imaged for the first time in the H-band at $\lambda=1.6$~$\mu$m by \cite{2015ApJ...807L...7C} using the \emph{Gemini Planet Imager} (\emph{GPI}). The observation revealed a very broad nearly edge-on disc with an observed width to mean radius ratio $\Delta r$/$r_{0} \sim$ 0.37, which makes the HD~115600 debris disc one of the broadest ever observed -- see Table ~\ref{Table1} for the main parameters of HD~115600. 

\cite{2015ApJ...807L...7C} used the radiative transfer code GRaTeR \citep{1999A&A...348..557A} to model the disc emission. They found that the disc, most likely primarily constituted of ice-icy/silicate grains, appears eccentric with $0.1<e<0.2$ and its projected centre offset by $x_{off}=0.018^{\arcsec}$ and $y_{off}=0.029^{\arcsec}$ compared to the stellar location leading to a total offset of $r=\sqrt{x_{off}^{2}+y_{off}^{2}}=0.034^{\arcsec}$ or $3.75$~AU. They proposed that such disc eccentricity and offset could have been sculpted by a potential planetary companion. The absence of a direct detection of such a companion by \emph{GPI} places an upper limit of the mass of the potential companion of $m_{p}<7$~$M_{J}$ if it is ideally located on an orbit outside the coronograph mask. The authors point out that their model, however, respectively overestimates (underestimates) the disc flux at the southeast (southwest) corner. Using predictions from the gap opening model developed by \cite{2015ApJ...798...83N}, as well as the planet-stirring scenario established by \cite{2009MNRAS.399.1403M}, only very loose constraints on the potential planetary companion could be derived and the authors concluded that either a superjovian planet orbiting at $a<30$AU or a super-Earth located at the very inner edge of the disc could sculpt the disc.

N-body simulations can be used to model the interaction between a debris disc and a planet. In these simulations, a disc, modeled by an ensemble of massless particles, orbits around a central body representing the star and feels the gravitational perturbation of another massive body. To test different planet configurations, a suite of these simulations must be run, and the planet parameter space (comprised of the planet mass and orbital elements) is traditionally explored by using a grid of set values \citep{2005ApJ...625..398D,2014A&A...563A..72F}. In this approach, the entire parameter space must be explored to isolate the best fit, and the precision of the best fit parameters directly depends on the size of the grid, and therefore a large number of simulations are required to reach high precision. Given the low number of currently resolved debris discs\footnote{The total resolved debris disc is 41 as referenced by http://www.circumstellardisks.org/ (accessed 25/07/16).}, N-body simulations have been so far used to study individual objects \citep{2014A&A...561A..43B,2015ApJ...815...61N}, however with the increased number of expected discoveries from the latest and next generation of instruments such as \emph{ALMA} and \emph{JWST}, a more efficient and statistical approach must be found. For example, the \textit{Monte Carlo Markov Chain} (MCMC) approach explores closely located points of the parameter space along a chain, and by accepting parameter space point resulting in better fits and rejecting point resulting in worse fits, the chain quickly converges toward the best fit region. This method has been used in astrophysics to estimate cosmological parameters with high precision \citep{2002PhRvD..66j3511L},  adjusting semi-analytic models of galaxy formation \citep{2013MNRAS.428.2001M} or fitting planetary orbital element from transit and radial velocity data \citep{2013PASP..125...83E}. \cite{2015NatCo...6E7599U} present the first study integrating N-body simulations within an \textit{MCMC} algorithm to estimate the pre-infall mass of the Carina galaxy before it joins the Milky Way satellites group. One focus of our study is to present a framework for testing if the \emph{MCMC} algorithm can be used to more efficiently probe the parameter space of N-body simulations of a planet shaping a debris disc than a traditional grid method.

In this paper, we investigate the role of planets in determining the morphology of the HD~115600 debris disc. We dynamically model the interaction between an exoplanet and the HD~115600 disc using our modified N-body integrator, and explore the parameter space of the planets' orbit and mass using both the sampling method \textit{MCMC} and a traditional grid method. To compare the results of our numerical simulations with observations of HD~115600, we create synthetic images from the simulations at a similar wavelength and resolution as the \emph{GPI} image. Thus this study has two primary aims to both test the \textit{MCMC} algorithm for exploring the parameter space in N-body simulations and also provide information on a potentially hidden companion that planet hunters can use in the future searches.

We first present our general method in Section 2. In order to derive the grain composition and properties required to turn our simulations into synthetic images, Section 3 presents our spectral modeling of the HD~115600 disc using the radiative transfer code \textit{MCFOST}. In Section 4, we introduce our numerical method for performing the dynamical simulations, before presenting our results in Section 5. A summary of our findings and conclusions are given in Section 6.

\begin{table}
\renewcommand{\arraystretch}{1.0}
\caption{Properties for HD~115600 from Currie et al. (2015).}
\label{Table1}
\centering
\begin{tabular}{cc}
\hline
 Stellar properties & \\
\hline
Spectral type & F2V/3V  \\
Age & 15 Myr$^{a}$\\
Luminosity $L_{\ast}$ & $\sim$ 4.8~$L_{\odot}^{b}$ \\
Mass $M_{\ast}$ &  $\sim$ $1.5~M_{\odot}$ \\
Distance & 110.5~pc$^{c}$ \\
\hline
 Disc properties & \\
\hline
Width $\Delta r$ &  37.5--55~AU \\
Mean radius $r_{0}$ & $ 48 \pm 1.1$~AU\\
$\Delta r$/$r_{0}$ & 0.37 \\
Eccentricity $e$ & 0.1--0.2 \\
Mass & 0.05 $M_{\rm moon}$\\
Proj. offset $\delta$ & $3.7 \pm 1.5$~AU \\
Line-of-sight inclination $i$ & $79.5 \pm 0.5^{\circ}$\\
PA & $24 \pm 0.5^{\circ}$\\
\hline
\end{tabular}
\\
\small{ $^{a}$ Pecaut et al. (2012), $^{b}$ Chen et al. (2011), $^{c}$ van Leeuwen (2007).}
\end{table}

\section{General method}
One of our goals in this work is to determine the parameters of a potential planetary companion sculpting the disc geometry as observed in the \emph{GPI} image of \cite{2015ApJ...807L...7C}. To achieve this, we run dynamical simulations of a planet interacting with a debris disc, then combine the results of these simulations with a radiative transfer code to create a synthetic image of the disc and compare the disc's geometric parameters extracted from the synthetic image to the observations. Modeling the image with a radiative transfer code, however, not only requires the disc density structure (which is provided by the dynamical simulations), but also requires a set of dust properties, such as the grain size distribution, which has not been constrained by the observations. 

Our modeling therefore consists in two parts: we first model the spectral energy distribution using a radiative transfer code to determine the additional dust properties required to model the image. We then combine our dynamical simulations covering a range of initial planetary parameters with the radiative transfer code using the derived dust properties in order to model the image.

\subsection{Step 1: SED modeling}
To model the spectral energy distribution, we use the radiative transfer code, \emph{MCFOST} \citep{2006A&A...459..797P}. To calculate the total disc flux at a specific wavelength \emph{MCFOST} needs (i) a disc density structure and (ii) information regarding the dust composition and optical properties. 

In this first step, the disc density structure is calculated by using a parametric model which assumes a radial power-law. The disc size is based on the inner and outer radius estimated by \cite{2015ApJ...807L...7C} and we explore different values for the total dust mass, $M_{d}$.  Additional details are presented in Section 3. To determine the dust composition and optical properties which best reproduce the observed SED, we explore a range of values for the unknown properties related to the dust grain size distribution, as well as the unknown ratio of silicates to water ice in the grain given the predicted dust composition by \cite{2015ApJ...807L...7C}.

Using the derived parametric disc structure and the dust properties, \emph{MCFOST} computes the total flux, $F_{i}$, for each wavelength, $\lambda_{i}$, corresponding to the data from the literature -- see Table \ref{Table2}. We use a $\chi^{2}$ method to compare the computed and the observed SED, and the best fit parameters for the dust composition is obtained by minimizing the $\chi^{2}$. Additional details, such the range of parameters explored and results of this modeling, are presented in Section 3. 

\subsection{Step 2: Image modeling}
Once the best fit to the dust properties are obtained, we can then move on to the image modeling. Here we again use \emph{MCFOST}, but now the disc structure is generated from the numerical simulations of a planet interacting with a debris disc. We cover a broad range of initial planetary masses, $m_{p}$, semi-major axes, $a_{p}$, and eccentricities, $e_{p}$ -- further details on the dynamical simulations are provided in Sections 4.1 \& 4.3.

The dust distribution resulting from the dynamical simulations is converted into a disc structure, which is described in Section 4.2. \emph{MCFOST} uses this disc structure and the dust properties derived in step 1 to compute a synthetic image at $\lambda=1.6~\mu$m, and we then extract the disc geometry parameters from the synthetic image. We use a $\chi^{2}$ method to compare the disc geometric parameters from the synthetic image with the observed parameters determined by \cite{2015ApJ...807L...7C}.

Once the $\chi^{2}$ of a model is calculated, we run a new dynamical simulation for another initial planetary configuration using a \emph{Monte Carlo Markov Chain} sampling method to explore the parameter space. In total, we explore more than 700 planetary configurations before determining the best fit parameters $m_{p}$, $a_{p}$, and $e_{p}$ and the resulting best fit disc geometric parameters. More information about this step is provided in Section 4.3.

\section{SED modeling}
\subsection{Description of the method}
\subsubsection{\emph{MCFOST} requirements}
To compare a dynamical model to the observations, the numerical simulation must be converted into a synthetic image with the radiative transfer code \emph{MCFOST}. \emph{MCFOST} tracks monochromatic photon packets propagating through a 3D spatial grid in order to derive the temperature structure of the disc. The radiative transfer code therefore requires a prior 3D disc density structure, as well as information regarding the dust composition and optical properties, which, in the case of for HD~115600, are as yet unconstrained by the observations. These required properties are the five following parameters: the minimal and maximal size of the grain size distribution, $s_{min}$ and  $s_{max}$, the power law exponent, $p$, of the classical grain size distribution \citep{1969JGR....74.2531D}, $dn(s) \propto s^{p} ds$, the total dust mass, $M_{d}$, and using the predicted composition of the dust by \cite{2015ApJ...807L...7C}, the ratio of silicate to icy component in the grain composition, $R_{silicate/ice}$. Therefore our first aim is determining these five free parameters.

To achieve this, we first assume a simplistic parametric model for the disc density structure, and by creating SEDs over a broad range of values for the five free parameters, we estimate their best fit values by matching the observed SED with multiwavelength photometry data -- see Table~\ref{Table2}. We use a parametric disc density structure of the following form: $\Sigma(r)=r^{\alpha}$ where $\alpha =-3.5$, with the inner and outer disc radius and total dust mass matching the observed values from Table~\ref{Table1}.

\subsubsection{Parameter space exploration}
 Using the information provided by \cite{2015ApJ...807L...7C}, we explore the parameter space of free parameters as follows: we first generate 8575 different sets of dust properties by exploring a 5D grid comprising 7 values for $s_{min}$ between $0.0001~\mu$m$<s_{min}<0.1~\mu$m in even logarithmic steps; 7 values of $s_{max}=25, 50, 75, 100, 200, 400$ and $600~\mu$m; 5 values for $p=-3.2, -3.5,-3.8,-4$ and $-4.2$; 5 values for $M_{d}=0.05, 0.075, 0.1, 0.15, 0.2~M_{\rm moon}$; and 7 values for $R_{silicate/ice}$ between $20\%<R_{silicate/ice}<80\%$ in even steps. In addition, two models for the grain mixture are explored: firstly grains made of a mixture of silicate and ice according the Bruggeman's model, and secondly grains with a silicate core and a coating of ice \citep{1989AdSpR...9....3G}. We explored each set of five dust parameters using both grain mixture models, which correspond to a total of 17150 different sets of dust properties.

We estimate the best fit for the grain parameters using a $\chi_{\rm SED}^{2}$ minimization method, where $\chi_{\rm SED}^{2}$ is defined as:
\begin{equation}
\chi_{\rm SED}^{2}=\frac{1}{N-df-1}\sum_{i}^{N} \frac{(F_{Obs/i} - F_{Sim/i})^{2}}{\sigma^{2}_{Obs/i}},
\end{equation}
where $F_{Obs/i}$ and $\sigma_{Obs/i}$ are respectively each $i$-th value of the $N$ observed fluxes and uncertainties from Table~\ref{Table2} at the wavelength $\lambda_{i}$, $df$ is the number of parameters to fit and $F_{Sim/i}$ is the synthetic flux derived by \emph{MCFOST} at the same wavelength. Given the absence of associated uncertainties for some photometric values in Table~\ref{Table2}, those values do not contribute to the calculation of the total $\chi_{\rm SED}^{2}$.
\begin{table}
\renewcommand{\arraystretch}{0.9}
\caption{Photometry of HD~115600 from different instruments.}
\label{Table2}
\centering
\begin{tabular}{cccc}
\hline
$\lambda$ ($\mu$m) & F (mJy) & $\sigma_{Obs}$ (mJy)& Reference\\
\hline
0.44 & 1310 &  &     {\cite{2011AJ....142...15G}}\\
0.482 & 1700 &  &   {\cite{2016yCat.2336....0H}}\\
0.55 & 1860 &    &    {\cite{2011AJ....142...15G}}\\
1.24 & 1620 & 30  &  {\cite{2010AJ....139.2440R}}\\
1.65 & 1170  & 40 &  {\cite{2010AJ....139.2440R}}\\
2.16 & 784 & 17 &   {\cite{2010AJ....139.2440R}}\\
3.35 & 374 & 9 &   {\cite{2012yCat.2311....0C}}\\
4.6 &  205 &  4 &   {\cite{2012yCat.2311....0C}}\\
8.6  & 73.6 & 6.9 &  {\cite{2010A&A...514A...1I}}\\
11.6 & 39.8 & 0.6 & {\cite{2012yCat.2311....0C}}\\
13.0 & 44.63 & 1.48  &  \cite{2014ApJS..211...25C}\\
22.1 & 12.3 &  2 &  \cite{2012yCat.2311....0C}\\
24.0 & 120.23 & 2.41  &   \cite{2014ApJS..211...25C}\\
31.0 & 198.38 & 18.35 &   \cite{2014ApJS..211...25C}\\
70.0 & 180.0 &  21.93 &  \cite{2014ApJS..211...25C}\\
\hline
\end{tabular}
\end{table}

\subsection{Results}
\begin{figure*}
\begin{center}
 \subfloat{\includegraphics[width=65mm,height=48mm]{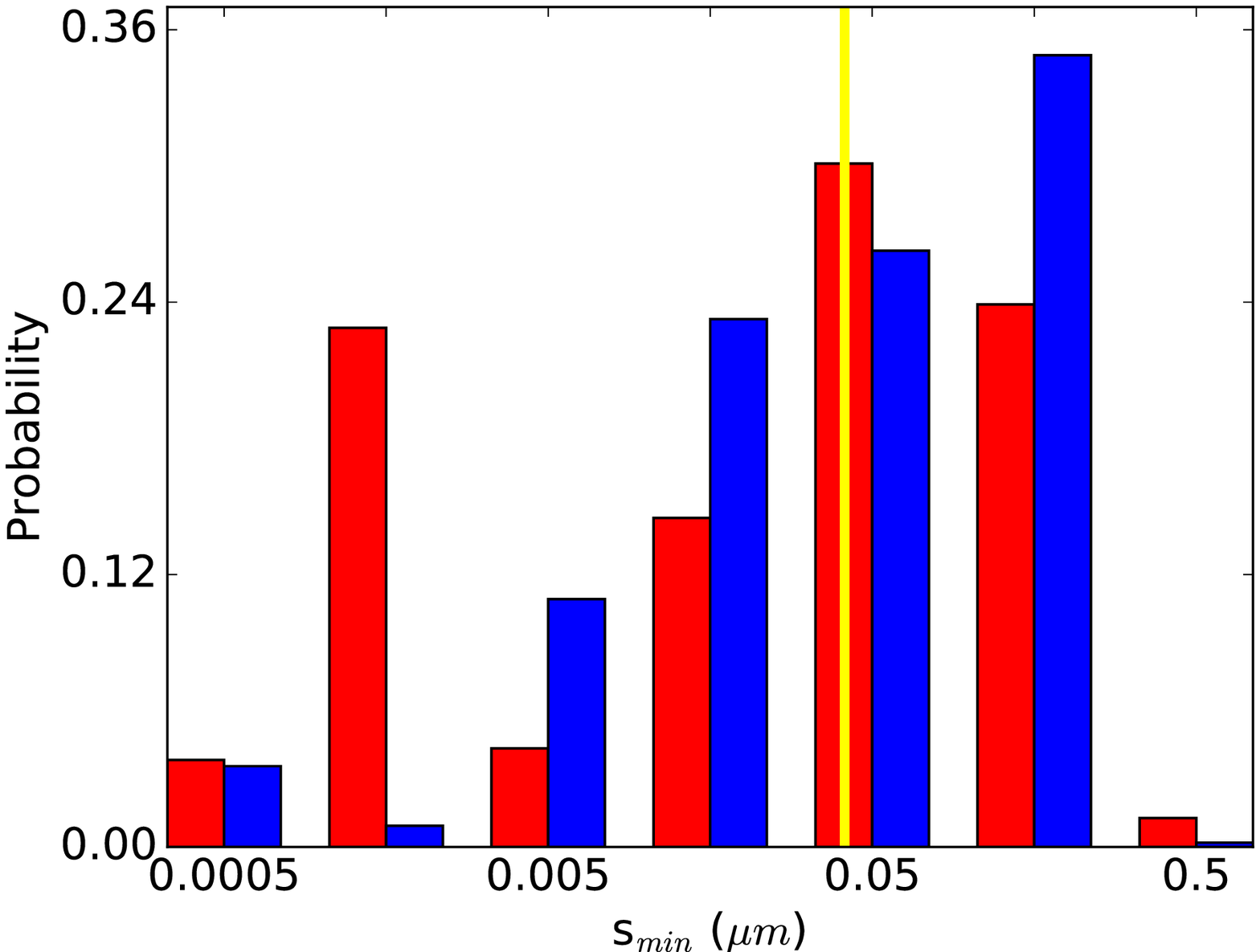}}  
 \subfloat{\includegraphics[width=65mm,height=48mm]{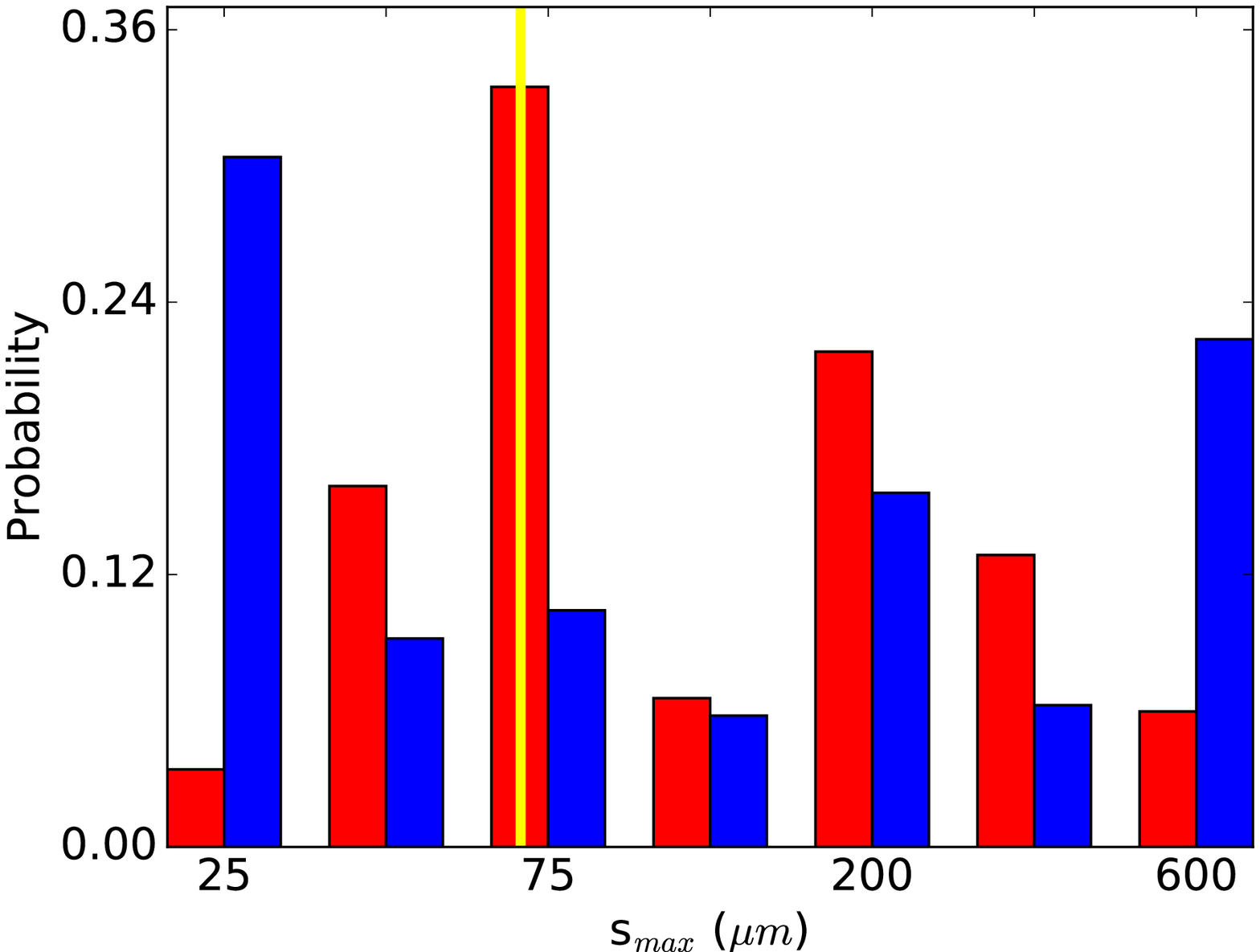}}       
 \subfloat{\includegraphics[width=65mm,height=48mm]{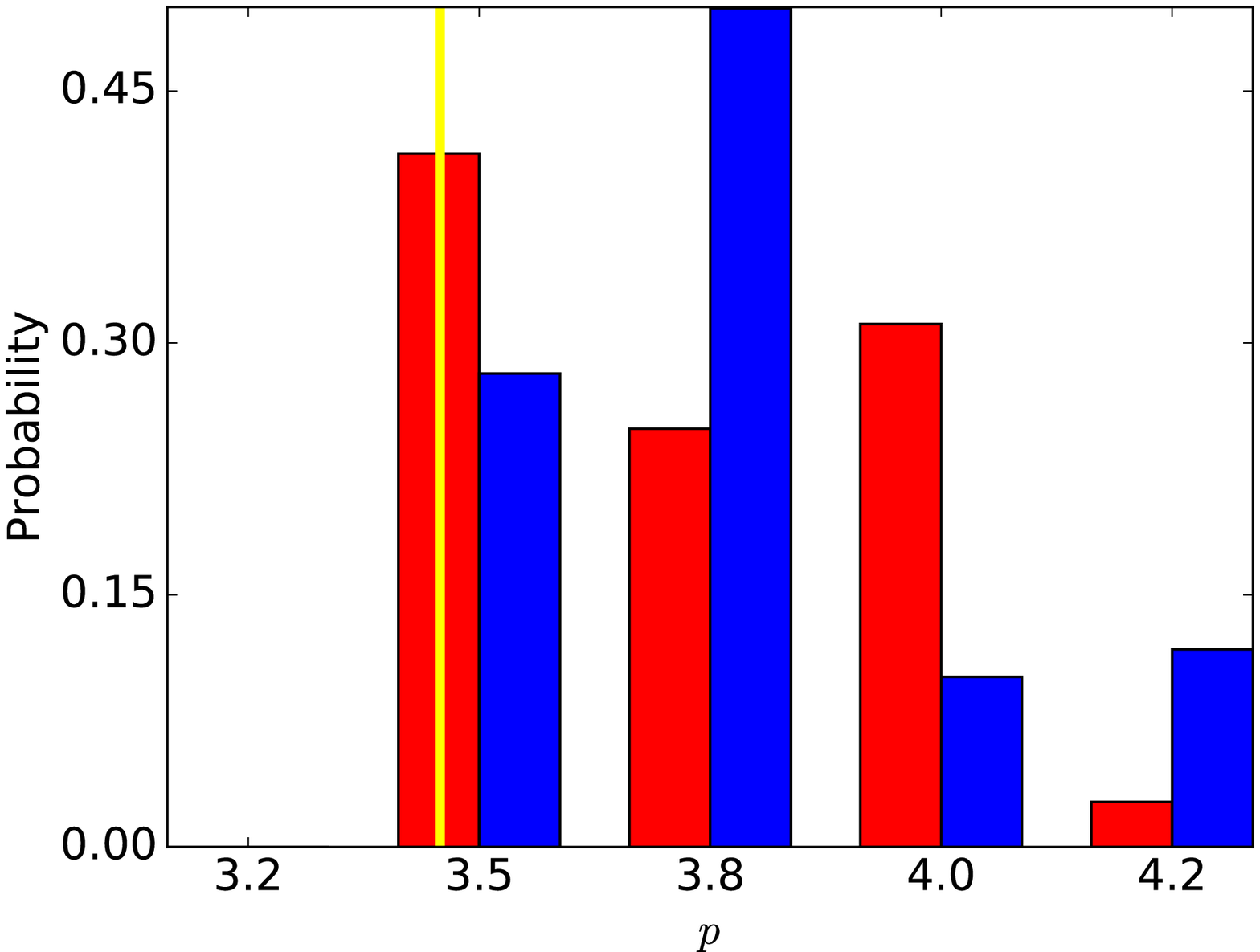}} \\
 \subfloat{\includegraphics[width=65mm,height=48mm]{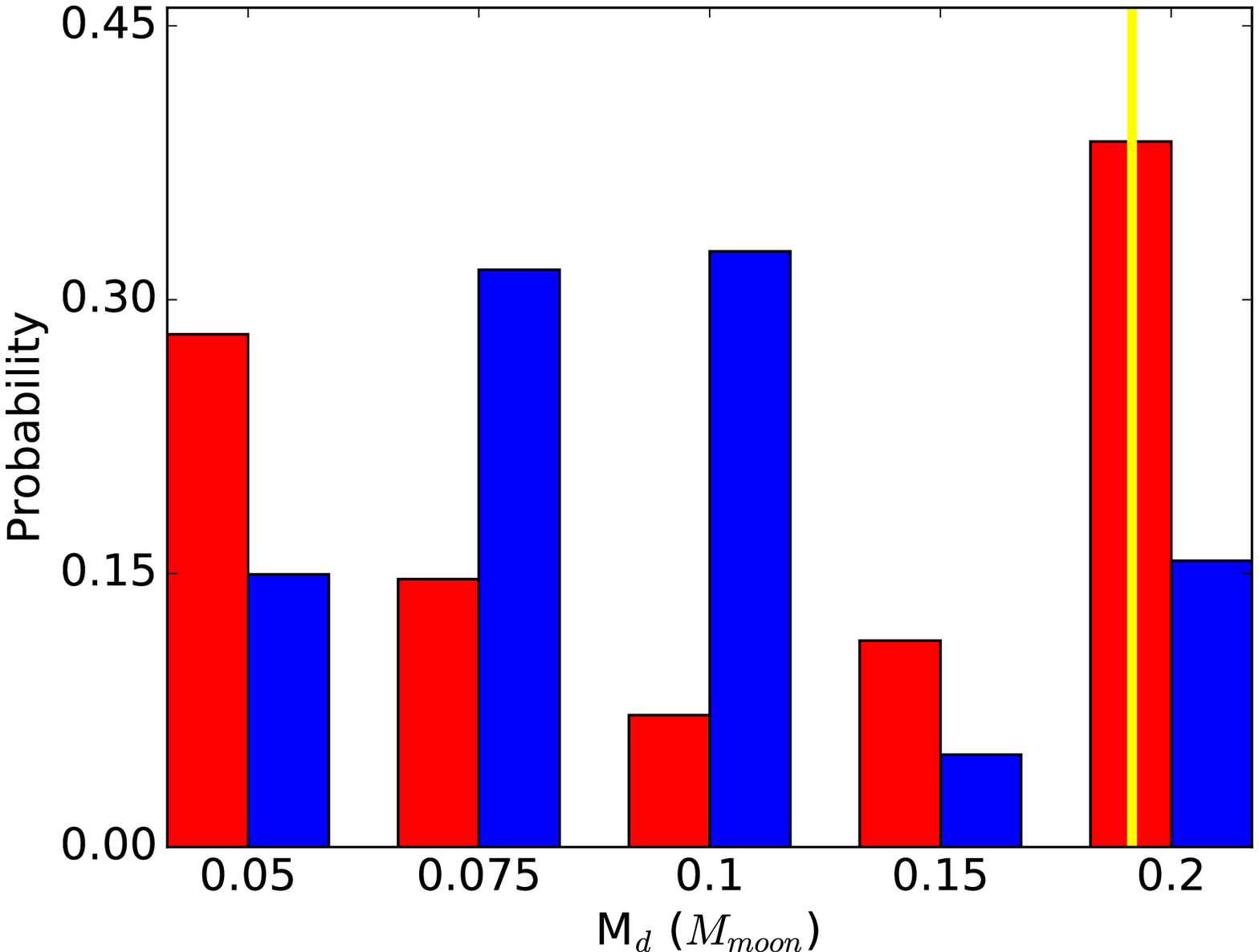}} 
 \subfloat{\includegraphics[width=65mm,height=48mm]{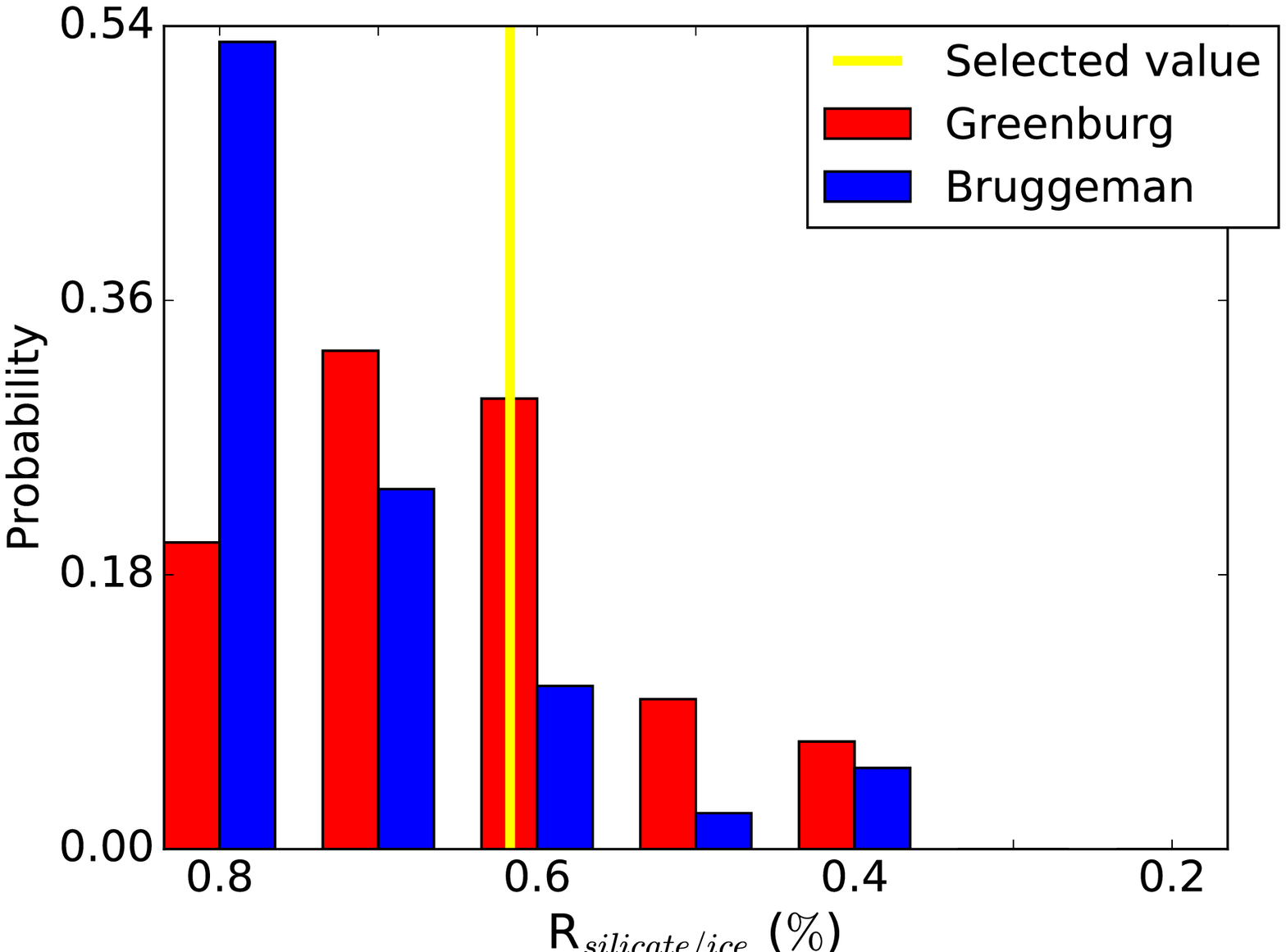}}                         
 \caption{Results from SED modeling: probability distribution for the minimal and maximal grain size, $s_{min}$, $s_{max}$, the silicate-to-ice ratio, $R_{silicate/ice}$, the power law exponent of the grain distribution, $p$, and the total dust mass $M_{d}$, according both the Greenburg and Bruggeman grain model.} 
\label{fig:1}  
\end{center}
\end{figure*}
The $\chi^{2}$ best fit for the Bruggeman grain model resulted in $\chi_{\rm SED}^{2}=43.1$  while the best fit with a Greenburg model led to $\chi_{\rm SED}^{2}=40.6$, slightly favoring a grain model based on the Greenburg model where the grain is made of a core of silicate with an external coating of ice. For the unknown dust parameters, the probability distribution for both grain models are presented in Figure~\ref{fig:1}. Both models seem to privilege a minimal grain size between $0.01< s_{min}< 0.1~\mu$m, a power-law index between $-3.8<p<-3.5$, as well as, a high ratio of silicate-to-ice composition between $0.6< R_{silicate/ice} <0.8$. On the other hand, very little constraint can be derived for the maximal grain size and the total dust mass of the disc, and the similarity between both $\chi^{2}$ prevents us from drawing a clear conclusion regarding the best fit grain model. 

We however decide to use a Greenburg model in the rest of this study as this model is, in addition to have the lower $\chi^{2}$, in good agreement with the grains' reflectance spectra from Currie et al. which favours a water ice composition. To later create synthetic image and spectral energy distribution using a disc density structure resulting from dynamical simulations, we therefore use the parameters for the Greenburg best fit model ($\chi_{\rm SED}^{2}=40.6$): a minimal and maximal grain size of $s_{min}=0.05~\mu$m and $s_{max}=75~\mu$m using a grain size exponent of $p=-3.5$, a total dust mass of $M_{d}=0.2~M_{\rm Moon}$, and a grain composition made of $60\%$ silicate (core) with $40\%$ water ice component (mantle). We note that the best fit value for each parameter corresponds to the peak of its Greenburg model probability distribution, with the exception of $R_{silicate/ice}$, for which the distribution peaks at 0.7.

\section{Image modeling}
Now that we have the best fit for the dust properties, we aim to reproduce the observed disc geometry resulting from the planet shaping the disc. In this section, we introduce the numerical integrator used to model the dynamical interaction the planet and dusty debris disc, and explain how we use the simulation results as input for the radiative transfer code. We then describe how the disc geometry parameters, such as projected offset, disc eccentricity, peak brightness location and width, are determined from the synthetic images and used to compare with the observations.
\subsection{Dynamical simulations}
\subsubsection{Simulations set-up}
To dynamically model the interaction between the debris disc and a potential planetary companion, we use our modified version for the N-body code \emph{SWIFT} \citep{1994Icar..108...18L} which includes radiation forces (radiation pressure and Poynting Robertson) and stellar wind acting on small grains  -- see \cite{2015PASA...32...39T} for a complete description of the numerical method. The disc orbits a central star of mass $M=1.5~$M$_{\odot}$. We choose to model the debris disc by an initially dynamically warm disc, an appropriate initial assumption for systems assumed to host a several Jupiter masses planet \citep{2015PASA...32...39T}. A set of $N=$5500 massless test particles represent the grains and are initially located between $37.5<a<55$~AU, with $0<e<0.2$ and $i<1^{\circ}$ -- see the left column of Table \ref{Table3}. 

Near-infrared images, such as the \emph{GPI} image, trace the disc thermal emission resulting from small dust grains. Due to their small size, the grain orbits are sensitive to stellar radiation forces: while the radial component of the radiation pressure tends to reduce the stellar gravitational attraction, its tangential component (the Poynting-Robertson effect) tends to drag dust grains inwards. Similarly, stellar winds accelerate their decay toward the star via the impact of protons on dust grains. Therefore, while the larger grains in the disc are expected to remain close to their parent planetesimal belt, smaller grains are expected to have their orbits perturbed by stellar radiation, and an accurate dynamical model must be taking these forces into account. Since the blowout grain size in HD~115600 has been estimated to be $s_{\rm blow}= 3~\mu$m  \citep{2015ApJ...807L...7C}, we model grains with a size of $s=8~\mu$m with the corresponding ratio between the gravitational and radiation forces is $\beta=0.2$. The stellar wind coefficient is set to $sw=0.6$, corresponding to stellar wind in the solar system felt by icy grains \citep{1982A&A...107...97M} and appropriate for the best fit dust grain model determined in Section 3.2.

For the planetary companion, we cover a range of initial masses, $m_{p}$, semi-major axes, $a_{p}$ and eccentricities, $e_{p}$ using a \emph{Monte Carlo Markov Chain} (MCMC) sampling method, with each set of parameters representing an unique model. See Section 4.3 for more details.

\subsubsection{Constraining the planetary mass}
\begin{figure}
\begin{center}
  \includegraphics[width=90mm,height=67mm]{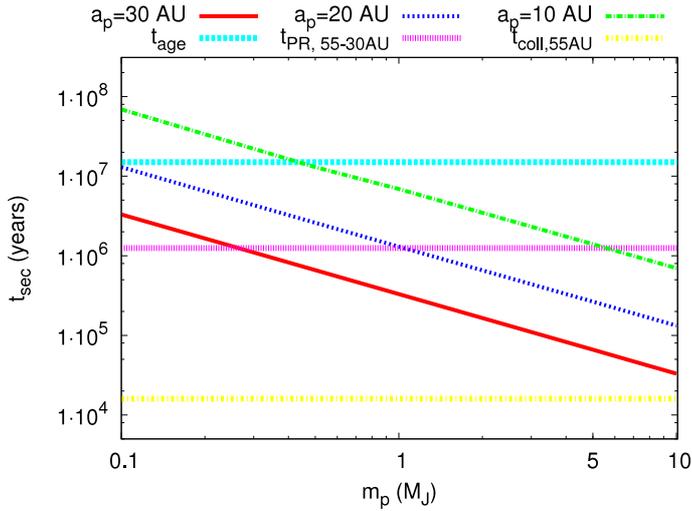}                            
  \caption{Secular timescale, $t_{sec}$, as a function of the planetary mass, $m_{p}$, for a planet respectively located at $a_{p}=$10, 20 and 30~AU around a 1.5~M$_{\odot}$. Overplotted are the age of the system, $t_{age}$, the collisional timescale, $t_{\rm coll, 55 AU}$, for a grain at the outer edge of disc ($a=55$~AU), as well as the Poynting-Robertson timescale for a grain at $a=55$~AU to reach the critical zone at proximity of the planet ($a\sim 30$~AU), $t_{\rm PR, 55-30 AU}$.} 
\label{fig:2}  
\end{center}
\end{figure}

The young age of HD~115600 ($t_{age}=15$~Myr) can provide an additional constraint on the potential planet's mass. If we first assume that the planet is forcing the disc eccentricity to precess around a forced value via secular interactions, then the system must be at least older than the secular interaction timescale, $t_{sec} < t_{age}$. This timescale represents the precession timescale of the disc eccentricity around the value imposed by the planet -- see \cite{1999ApJ...527..918W} for a complete definition -- and the disc is expected to settle in its final secularly forced configuration after a period of a few $t_{sec}$ \citep{2014MNRAS.443.2541P}. Figure \ref{fig:2} shows the secular timescale, $t_{sec}$, as a function of the planetary mass, $m_{p}$, for a planet respectively located at $a_{p}=$10, 20 and 30~AU around a 1.5~M$_{\odot}$ star.. Under this assumption, as seen in Figure \ref{fig:2} only a planet more massive than 0.5~M$_{\rm J}$ located at $a_{p}=10$~AU matches this criterion, while if the planet is located beyond 20~AU, then its mass must be at least 0.1~M$_{\rm J}$. 

Another constraint is that the planet must sculpt the disc before the small grains imaged in scattered light are destroyed. For older debris discs ($ > 100~$Myr), a steady-state regime, where new dust is continuously replenished by collisions of meter-sized planetesimals, is thought to be responsible for the dust survival on timescale longer than the Poynting Robertson timescale, the dust survival time before being removed by PR drag. In such discs, collisions are therefore very active and a dust grain of a specific size can only remain intact for a limited amount of time before undergoing a potentially destructive collision. The collisional timescale, which defines the period between two collisions for a grain at a specific stellar distance, at the edge of the outer disc is $t_{coll,55AU}= P_{55AU}/4\pi\tau$, where $P_{55 AU}$ is the Keplerian period at $55$~AU and $\tau$ is the effective optical depth. Using the estimated fractional luminosity of HD~115600 ($L_{\rm IR}/L_{\star}=1.7\times 10^{-3}$) as a proxy for the disc effective optical depth, we estimate the collisional timescale to be $t_{coll}=16000$~yrs. For the planet to shape the disc, the secular interaction must act on timescale smaller or similar to the collisional timescale: from Figure \ref{fig:2}, it is clear that only very massive ($>5-10~M_{\rm J}$) planets located close the inner edge of the disc ($a_{p}=30$~AU) could fulfill such requirement. No planet larger than $7~M_{\rm J}$, however, was detected by \emph{GPI}. 

For a disc as young as HD~115600, and in the absence of millimeter images revealing the presence of large grains in the disc and its underlying planetesimals population, it remains unclear if the micron-sized dust observed on the \emph{GPI} image results from a new generation created by planetesimal collisions or if the dust is a direct remnant of protoplanetary dust. If only dust grains of similar small sizes subsist from the protoplanetary disc phase, then, in the absence of dust replenishment by planetesimal collisions, Poynting-Roberston drag becomes the limiting factor of the long-term disc survival. Small dust grains still undergo collisions in the disc until the entire dust population has been removed by PR drag. In this case, another constraint for HD~115600 system is that the planet must sculpt the disc before all small grains are removed by the stellar radiation forces. We estimate the Poynting-Robertson timescale for a $8~\mu$m sized grain initially located at the outer edge of disc ($a=55$~AU) to be dragged into the proximity of the planet region ($a\sim 30$~AU) is $t_{\rm PR, 55-30 AU}= 400\times(55^{2}-30^{2})/\beta= 1.25$~Myr. This timescale is crucial, as grains undergoing a close encounter with a massive planet would be rapidly scattered away, and therefore an additional constraint is $t_{sec} < t_{\rm PR, 55-30 AU}$. Figure \ref{fig:2} indicates that for this constraint to be satisfied, a planet located at $a_{p}=30$~AU must be at least $m_{p}= 0.3~M_{J}$. If $a_{p}=20$~AU, then the Poynting-Robertson timescale for a grain to reach the planet zone from the disc outer edge increases ($t_{\rm PR, 55-20 AU} = 2.4~$Myr), and therefore a planet larger than $m_{p}=0.6~$M$_{\rm J}$ would be required to satisfy $t_{sec}<t_{\rm PR, 55-20 AU}$. Similarly a planet with a mass of at least $m_{p}= 2.0$~M$_{\rm J}$ at $a_{p}=10$~AU can still excite the disc eccentricity within the Poynting-Robertson timescale, $t_{\rm PR, 55-10 AU}\sim 4$~Myr.

Overall, this analysis indicates that if we assume secular interactions to be the key element in explaining the morphology of HD~115600, then the planet is certainly more likely to be a Jovian or super-Jovian ($0.5<m_{p}/M_{\rm J}<7+$) than a super-Earth planet, and therefore we will focus our exploration of the space parameter of the planetary mass within this range.

\subsubsection{Simulations duration and recording process}
For this study, we run the simulations for a duration $t_{sim}=t_{\rm PR, 55-30 AU}$, corresponding to roughly $0.1~t_{age}$ of HD~115600, which is the timescale for grains initially at the outer edge of the disc ($a=55~$AU) to be dragged into the proximity of the potential planet around $a \sim 30~$AU. The particle positions and velocities are recorded every $t_{record}=K \times P_{pl}$, where $P_{pl}$ the planetary orbit period, and $K$ is the number of sampling times for all particles, defined as $K=N\times t_{sim}/(5\times 10^{6} \times P_{pl})$. This number of sampling times ensures that we record the particle distribution at the same planetary phase and that the total number of records do not exceed $5\times 10^{6}$.

\subsection{Creating synthetic images}
To compare our simulation results with the observed H band image from Currie et al. (2015), we again use \textit{MCFOST} to create synthetic images at $\lambda=1.6~\mu$m. \emph{MCFOST} first derives the temperature structure of the disc before creating the synthetic image via a ray-tracing method.  We use the dust optical properties determined in Section 3.2 as input for the procedure. For the density structure of the disc, the total particle distribution recorded at every $t_{record}$ during the simulations is converted into a density distribution, assuming that the total mass of the disc is 0.2 M$_{\rm moon}$ as derived by our SED modelling in Section 3.2. Using the entire test particle record allows us to mimic the effect of continuous dust replenishment in the disc by collisions over the lifetime of the simulation -- see Appendix A for further discussion.This density distribution is then binned into a 3D cylindrical grid with 90 radial bins between $10<r<70$~AU, 120 azimuthal bins and 30 vertical bins between $z \pm 20$~AU.

We determine the disc geometry parameters to compare with the observations in Table~\ref{Table1} using a similar method as introduced in \cite{2016MNRAS.457.1690T}. To measure the projected disc offset, $\delta$, we fit an ellipse to the brightest pixel of each 120 azimuthal bins of the projected disc image to obtain the offset coordinates. For the radial location of the peak brightness, $r_0$, and the disc width ratio, $\Delta r$/$r_{0}$, we plot the radial surface brightness profile after azimuthally averaging the surface brightness in each of the 90 radial bins of the deprojected image. The location of the peak brightness, $r_{0}$, corresponds to the maximal surface brightness, while the disc width ratio, $\Delta r$/$r_{0}$, is defined as the FWHM of surface brightness profile, $\Delta r$, divided by the location of the peak brightness, $r_{0}$. The disc eccentricity is obtained by measuring the disc deprojected offset, $\delta_{deproj}$, which is obtained by fitting an ellipse to the brightest pixel of each 120 azimuthal bins of the deprojected disc, and assuming $e=\delta_{deproj}/r_{0}$.

It should be noted Currie et al. (2015) use their \emph{GPI} projected PSF-subtracted, wavelength collapsed image of HD~115600 to fit an ellipse to the disc from which they extract the observed mean radius, $r_{0}$, and projected offset, $\delta$. They then model the disc emission with a radiative transfer code and use the best fit model to constrain a (model-derived) eccentricity, $e$, and (deconvolved model-derived) disc width, $\Delta r/r_{0}$. In comparison, we create a synthetic image from our simulations, which has the same wavelength and pixel resolution as the \emph{GPI} $1.6~\mu$m image, but which is not convolved. We then use this synthetic deprojected deconvolved azimuthally-averaged disc image to extract the mean radius, $r_{0}$, and disc width, $\Delta r/r_{0}$. With the projected deconvolved synthetic image we fit an ellipse (using the same package as Currie et al.) to get the projected offset, $\delta$, and eccentricity, $e$.  Therefore, whereas Currie et al.'s mean radius and offset are extracted from their projected and PSF-substracted image, both our disc width and eccentricity values are extracted from a deconvolved model-fitting image. Despite these slightly different approaches, we use the parameters from Currie et al. (2015) presented in Table~\ref{Table1} to directly compare with the geometric disc parameters derived from our \emph{MCFOST} synthetic images.

To achieve this, we compute the $\chi_{\rm img}^{2}$ of the model defined as:
\begin{equation}
\chi_{\rm img}^{2}=\sum_{X} \frac{(X_{Obs} - X_{Sim})^{2}}{\sigma^{2}_{X/Obs} + \sigma^{2}_{X/Sim}}
\label{eq1}
\end{equation}
where $X$ is the parameter ($r_{0}$, $e$, $\Delta r/r_{0}$ and $\delta$), $\sigma$ is the standard deviation of the parameter being fit, and the subscripts \textit{Obs}, \textit{Sim} correspond to the \emph{GPI} observation and the simulation respectively. For the standard deviation of the observed parameters, $\sigma_{X/Obs}$, we use the values provided by \cite{2015ApJ...807L...7C} in Table \ref{Table1}. We statistically obtained the standard deviation of the simulated parameters, $\sigma_{X/Sim}$ using a bootstrap method -- see \cite{2016MNRAS.457.1690T} for more details. The final best fit model is therefore determined by minimizing $\chi_{\rm img}^{2}$.

\subsection{MCMC algorithm}
\subsubsection{The algorithm}
Determining the planetary orbit responsible for shaping the observed disc of HD~115600 requires the exploration of a 3D parameter space over $m_{p}$, $a_{p}$ and $e_{p}$. The traditional way to explore this parameter space is a brute force method, where the parameter space is represented by a grid over which each parameter is sampled in bins of a fixed width ($dm_{p}$, $da_{p}$ and $de_{p}$), with each value of the grid explored. In the literature, the typical number of simulations used to explore such a grid rarely exceed a few hundreds. This method is not only computationally expensive but also wasteful, since the entire parameter space needs to be explored whether the explored region is in proximity to the best fit region or not.

Here we use the more flexible \emph{MCMC} sampling method, which makes use of random numbers to create a chain from an initial set of parameters. Each set represents one point in the 3D parameter space, and the \emph{MCMC} explores the parameter space by moving from one point to another along a chain \citep{2011AAS...21740704A}. This process allows us to efficiently analyze the 3D parameter space to assess the target probability distribution functions associated with each parameter.  Such an algorithm requires: (i) an initial set of parameters ($m_{p1}$, $a_{p1}$ and $e_{p1}$) which is the starting point of the chain in the 3D parameter space being explored, (ii) a prior distribution of parameters from which $i$ additional sets of  parameters ($m_{pi}$, $a_{pi}$ and $e_{pi}$) are created for the chain to explore, and (iii) a transition operator indicating when the chain should jump from a set of parameters to the new set which becomes the centre of the prior distribution at the next iteration of the chain. 

Using the initial set of parameters,  $m_{p1}$, $a_{p1}$ and $e_{p1}$ (\textit{set 1}),  as the initial planet parameters along with the initial disc parameters from Table \ref{Table3}, the dynamical simulation is performed with our modified N-body integrator and the synthetic image is then created with \emph{MCFOST}. After extracting the geometric parameters of the disc from the synthetic image ($\delta$, $r_{0}$, $\Delta r/r_{0}$ and $e$), the $\chi^{2}_{\rm set 1}$ of the model is computed using Eq. \ref{eq1}. For each parameter, the initial prior distribution corresponds to a normal distribution centred on the initial values $m_{p1}$, $a_{p1}$ and $e_{p1}$, with a standard deviation, $\sigma_{a_{p}}$, $\sigma_{e_{p}}$ and $\sigma_{m_{p}}$, which has been tuned to allow the transition operator to sample the parameter space efficiently -- see Appendix A. A new set (\textit{set 2}) of parameters, $m_{p2}$, $a_{p2}$ and $e_{p2}$, is then created by drawing a value for each parameter from the prior distribution.  Now using \textit{set 2} as the initial planetary parameters, a new dynamical simulation is run and a new synthetic image is created, from which the $\chi^{2}_{\rm set 2}$ of the model is determined. 

 After this step, the transition operator will decide if the chain should jump and use \textit{set 2} instead of \textit{set 1} as the centre of the prior distribution to do the next sampling or not. The transition operator (also known as acceptance rate), $\alpha$, between two sets of initial parameters, \textit{set 1} and \textit{set 2}, of the chain is defined as:
\begin{equation}
\alpha=\exp(-\chi^{2}_{\rm set 2}+\chi^{2}_{\rm set 1})
\label{eq2}
\end{equation}
where $\chi^{2}$ is defined by Eq (\ref{eq1}). The condition required for the chain to make the jump is: $\alpha > \gamma$, where $\gamma$ is a real number drawn from the uniform distribution between 0 and 1, $U(0,1)$. If the jump is made, then the \textit{set 2} of parameters ($m_{p2}$, $a_{p2}$ and $e_{p2}$) will be used as the centre of the prior distribution to create a new set of parameters (\textit{set 3}). If not, a new \textit{set 2} will be drawn from the prior distribution centred on \textit{set 1}. This process allows the transition operator to guide the chain toward the best fit region of the parameter space.

\subsubsection{Optimization}
To estimate the target probability distribution for each parameter, a typical \emph{MCMC} chain is made of $10^{4}-10^{5}$ iterations when used to explore the parameter space for quick operations such as curve regressions. However, in this study, we focus on assessing the efficiency of using \emph{MCMC} rather than using a grid method to sample the parameter space of a suite of dynamical simulations coupled with a radiative transfer code, which is a complex and computationally expensive operation. Even though conducting an exploration with a grid method has the technical advantage that it can potentially be parallelized, most studies in the literature do not exceed a few hundreds grid points. We therefore use a limited number of 720 iterations in our \emph{MCMC} chain, comparable to the number of grid points traditionally used in other studies in order to compare the results of both methods using a limited sample of simulations. Consequences of this choice will be discussed in the next section. 

Predetermining the number of required iterations in the chain to reach convergence towards the best fit region requires an initial trial \emph{MCMC} chain and complex statistics to generally overestimate the number of necessary iterations \citep{Raftery95thenumber}. In the \emph{MCMC} literature -- see \cite{1996CowlesCarlin} or \cite{ETS2:ETS201899} for a complete review of \emph{MCMC} convergence tests, most tests use the output of the chain to diagnose convergence and we use the same procedure in this study. In addition, while the \emph{MCMC} chain is expected to converge regardless of the choice of the starting point of the chain (the first set of parameters $m_{p1}$, $a_{p1}$ and $e_{p1}$), the number of iterations leading to the convergence will depend on the initial starting point. In practice, a small number of iterations at the beginning of the chain are throw out : this aims to reduce the dependence of the target probability distribution of each parameter on the exact starting point. We plot the evolution of parameters within the chain as well as the autocorrelation function, which estimates how sets of parameters at different iterations in the chain are related, to determine a satisfactory number of iterations to remove as a ``burn-in period'' and reduce the impact of the starting point choice.

Three conditions should be checked when employing the \emph{MCMC} method: first, we have to ensure that the correct region of the parameter space has been explored. We check this requirement by running multiple chains with different starting values ($m_{p1}$, $a_{p1}$ and $e_{p1}$) to ensure that all chains converge towards a similar target probability distribution of the three parameters by conducting an analysis of variance of all target probability distributions. Secondly, to ensure the best fit has been found, the chain has to reach a stable state and further sampling of the parameter space should not significantly change the resulting target probability distribution of each parameters.  We check this criteria by plotting the evolution of each parameter to verify that the parameters have effectively converged toward the best fit values.

Finally, to ensure the best performance of the chain, the optimal acceptance rate, $\alpha$, should be between 25-40\% \citep{1998Robertsetal}, although an acceptance rate of 15\% to 50\% has been proven to be 80\% efficient by \cite{2001RobertsandRosenthal}. An acceptance rate which is too high indicates that the chain is not well-mixed, meaning that the sets of initial parameters are drawn too close to each other in the 3D parameter space and that the chain is not exploring the parameter space fast enough. On the other hand, too small an acceptance rate means that too many candidates are rejected and that the chain is not moving through the parameter space efficiently.
We fulfill the acceptance rate condition by adjusting the standard deviation for each parameter ($\sigma_{a_{p}}=0.90$~AU, $\sigma_{e_{p}}=0.02$ and $\sigma_{m_{p}}=0.25~M_{\rm J}$) of the normal prior distribution from which the initial sets of parameters in the chain are randomly drawn -- see Appendix B.  We can then marginalize each parameter to obtain its probability distribution. 

\begin{table}
\renewcommand{\arraystretch}{1.0}
\caption{Summary of initial conditions of the test particles used in all simulations and of the planet parameters explored over the multiple MCMC chains.}
\label{Table3}
\centering
\begin{tabular}{ccc}
\hline
Parameters &  Particles & Planet \\
\hline
Semi-major axis (AU)  &  $37.5<a<55$  & $16.1<a_{p}<34.5$\\
Eccentricity  &  $0<e<0.2$ & $0.008<e_{p}<0.36$\\
Inclination  & $0<i<1^{\circ}$  & $i_{p}=0^{\circ}$\\
Mass ($M_{\rm J}$) & massless & $1.25<m_{p}<9.48$\\
\hline
\end{tabular}
\end{table}

\section{Results}
\subsection{Convergence of the chain}
\begin{figure*}
\begin{center}
  \includegraphics[width=160mm,height=118mm]{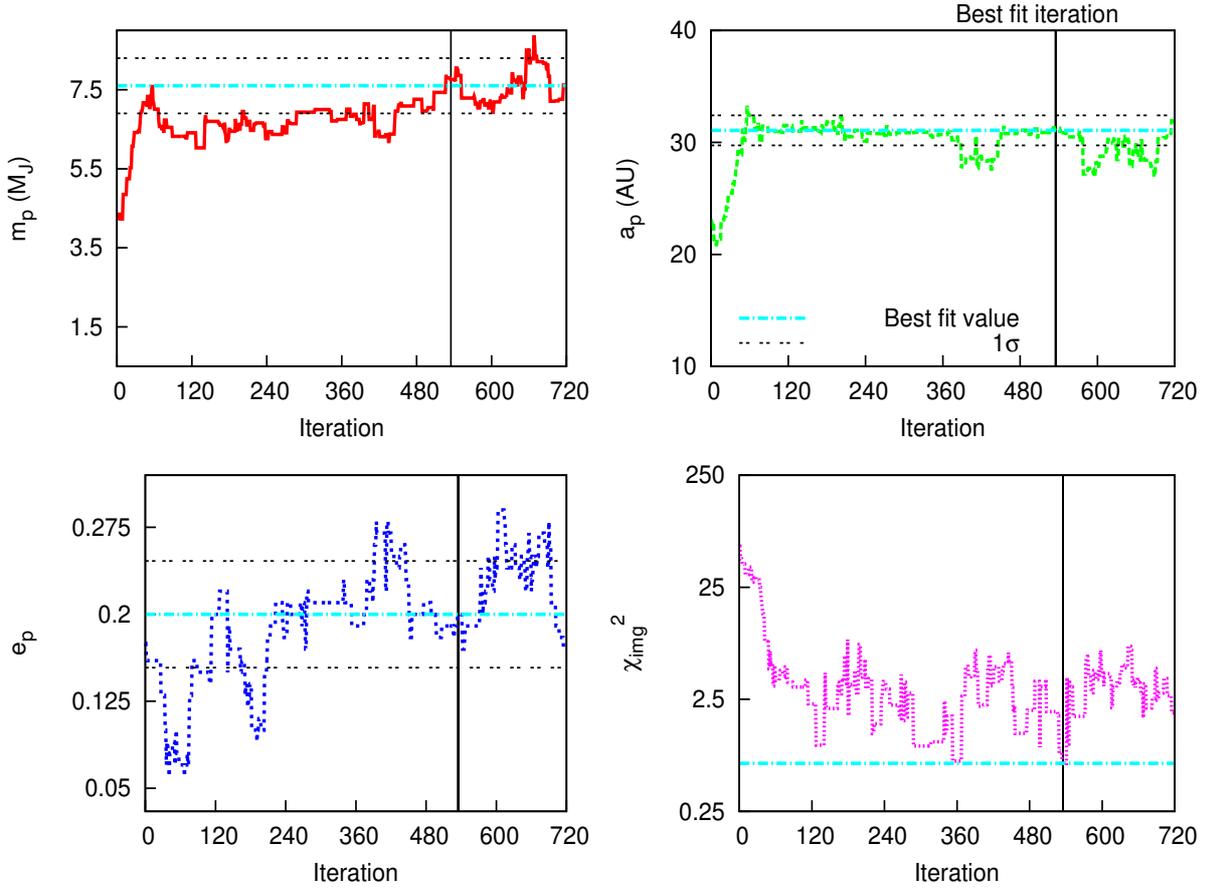}                            
  \caption{Traceplot of the 3 planetary parameters ($m_{p}, a_{p}, e_{p}$) and the evolution of the $\chi^{2}$ value. The blue dashed line represents the final best fit values, and the black dashed lines represent the 1$\sigma$ interval from Table 4. The vertical black line is the best fit iteration.} 
\label{fig:3}  
\end{center}
\end{figure*} 
We start the \emph{MCMC} chain 1 with $m_{p1}=4.25$~$M_{\rm J}$, $a_{p1}=22.5$~AU and $e_{p1}=0.175$, which roughly represents the centre of the 3D parameter space to be explored. This model resulted in $\chi^{2}_{img1}=58.2$ and we let the chain run for an additional 719 iterations. The final acceptance rate in the chain is $23.6\%$, which is in the optimal range. 

We then used 3 different tests to verify that the chain has converged towards the best fit region: we plot the evolution of the parameters to visualize the convergence, we use the autocorrelation to inspect the mixing level inside the chain, and we use an analysis of variance to ensure convergence has been reached statistically. The first simple test is to visually analyze the evolution of each parameter with increasing number of iterations through the chain. After the burn-in period, the chain should converge significantly toward the best fit parameters with no significant departures, which we quantify by the chain having reached a region within $1-1.5~\sigma$ of the best fit. We plot the evolution of each parameter in Figure \ref{fig:3}. After 100 iterations of the chain, models have reached $\chi^{2}_{img}<7.5$ and all the parameters in the chain have converged toward their best fit value (which will be presented in the next section) and only oscillate around these values within $\pm 1.5~\sigma$ until the end of the 720 iterations. To derive the probability distribution of each parameters, we therefore remove the first 100 iterations of the chain as a burn-in period in order to reduce the impact of choosing a starting point in the chain on the chain statistics. However, this method can not guarantee that the chain is not trapped in a local minima, and can not ensure that all relevant regions of the parameter space have been explored. 

\begin{figure}
\begin{center}
  \includegraphics[width=90mm,height=67mm]{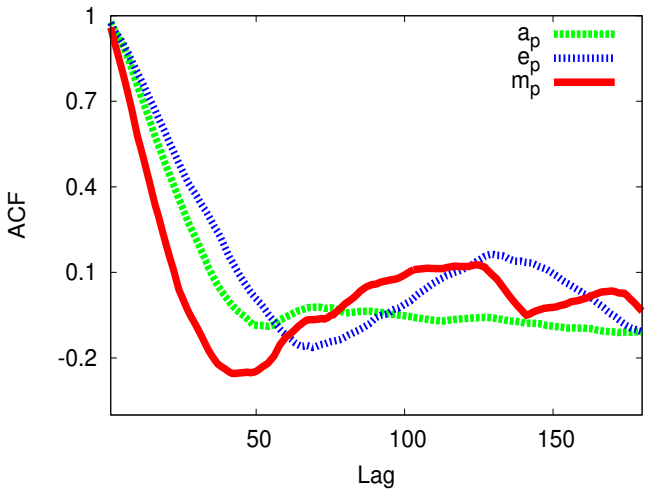}                            
  \caption{The autocorrelation function (ACF) against lag for the 3 planetary parameters.} 
\label{fig:4}  
\end{center}
\end{figure} 

Autocorrelation is used to quantify the dependency of the parameter sets at different iterations inside the chain. Autocorrelation is defined as:
\begin{equation}
\rho_{h}=\frac{\sum_{i=1}^{N-h} (\gamma_{i}-\bar{\gamma})(\gamma_{i+h}-\bar{\gamma})}{\sum_{i=1}^{N}(\gamma_{i}-\bar{\gamma})^{2}}
\end{equation}
where $\rho$ is the autocorrelation parameter, $N$ the number of iterations, $\gamma$ is one of the 3 parameters ($m_{p}$, $a_{p}$ and $e_{p}$) and $h$ is the lag \citep{1986MolPh..57...89S}. Good mixing in the chain is indicated by a decreasing correlation as the lag increases, meaning that the parameter values of two sets widely separated in time are independent of each other. We plot the autocorrelation for each parameter in Figure \ref{fig:4}, and after $\sim$ 100 iterations, the autocorrelation has dropped from 1.0 to less than $\pm$0.1 and the slope continues to converge toward 0 with increasing lag. This not only confirms a correct level of mixing in the chain, but also supports our choice of using the first 100 iterations as a burn-in period.

Another popular convergence test for \emph{MCMC} chains developed by \cite{1992GelmanRubin} consists of doing an analysis of variance over several chains with different starting values to ensure they all converge toward a similar target probability distribution. Finding the same target distribution by running multiple chains with different initial starting points in the parameter space ensures that the global minima is found. We follow this method by running two additional chains, starting at different initial set of values: we use $m_{p1}=2.62$~$M_{\rm J}$, $a_{p1}=28.75$~AU and $e_{p1}=0.237$ for chain 2, and $m_{p1}=5.875$~$M_{\rm J}$, $a_{p1}=16.25$~AU and $e_{p1}=0.11$ for chain 3. These values correspond to the $1/4$ and $3/4$ positions in the parameter space for each parameter. A summary of the range of planetary parameters explored over the 3 \emph{MCMC} chains is given in Table \ref{Table3}. We then compare each parameter variance within and between the chains to examine how the determination of the best fit value within a chain compares with the best fit values in the other chains. Using the formulation of \cite{1996CowlesCarlin}, the potential scale reduction factor (PSRF), given by $\sqrt{R}$, compares the between-chain variance, $B$, to the within-chain variance, $W$: 
\begin{equation}
\sqrt{R}=\sqrt{\frac{(1-\frac{1}{N})W+\frac{1}{N}B}{W}}
\end{equation}
where  $N$ the number of iterations and
\begin{eqnarray}
B &=& \frac{N}{M-1}\sum_{j=1}^{M}\left(\frac{1}{N}\sum_{i=1}^{N}\gamma_{i}^{(j)} - \frac{1}{M}\sum_{j=1}^{M}\frac{1}{N}\sum_{i=1}^{N}\gamma_{i}^{(j)} \right)^{2} ,\\
W &=& \frac{1}{M}\sum_{j=1}^{M}\frac{1}{N-1}\sum_{i=1}^{N} \left( \gamma_{i}^{(j)} -\frac{1}{N}\sum_{i=1}^{N}\gamma_{i}^{(j)} \right)^{2},
\end{eqnarray}
with $\gamma$ being one of the 3 parameters ($m_{p}$, $a_{p}$ and $e_{p}$) and $M$ the number of chains used. If each chain has converged towards a similar probability distribution, the PSRF value should be $1<\sqrt{R}<1.2$. Calculating the variance and PSRF for each parameters over the 3 chains after 720 iterations leads to: $\sqrt{R_{a_{p}}}=1.00$, $\sqrt{R_{e_{p}}}=1.10$ and $\sqrt{R_{m_{p}}}=1.35$. Although running additional iterations in the chains could furthermore reduce the PSRF of the planetary mass, $m_{p}$, the current values still indicate that all chains are converging toward best-fit models with similar $a_{p}$ and $e_{p}$.

\subsection{The best fit model}
\begin{figure*}
\begin{center}
\includegraphics[width=160mm,height=160mm]{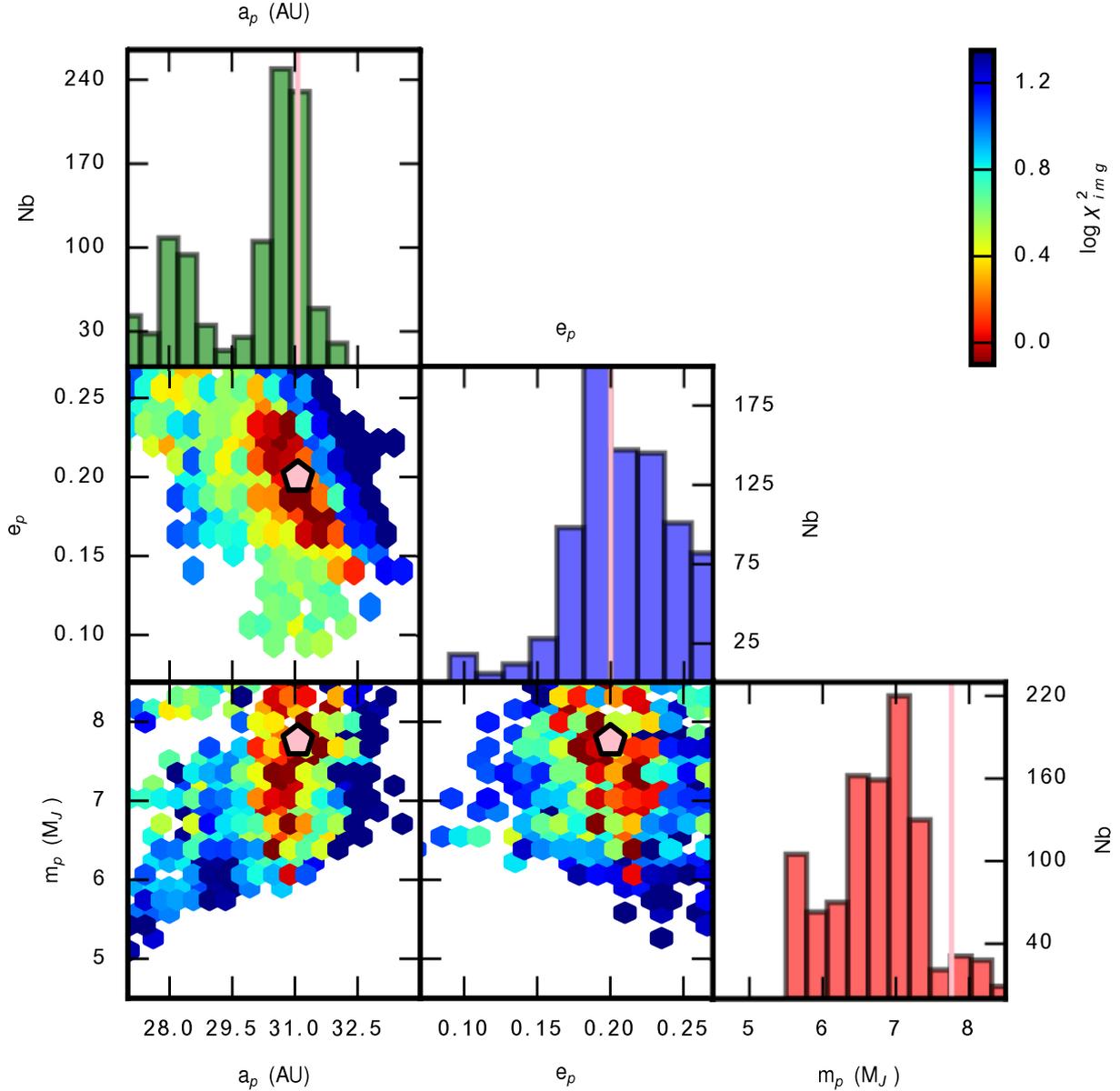}
  \caption{The top panels of each row presents the Bayesian probability distribution for each planetary parameter with the best-fit values indicated by the pink line. The other panels show a $\chi^{2}$ colourmap for each parameter pair, where the pink hexagon corresponds to the best-fit value.} 
\label{fig:5}  
\end{center}
\end{figure*} 
Figure \ref{fig:5} presents the results of the run of the \emph{MCMC} chain 1 over 720 iterations, for which the burn-in period of 100 iterations was removed. The top panel of each column represents the Bayesian probability distribution of each parameter, which is directly obtained by plotting the frequency of the values taken by each parameter in the chain, while the vertical pink lines corresponding to the best fit value. The best fit model ($\chi_{img}^{2}=0.67$) was found for a planet with $m_{p}=7.76~M_{\rm J}$, $a_{p}=31.07$~AU and $e_{p}=0.20$, with the planet semi-major axis and eccentricity best fit values corresponding to the peak of the probability distribution. Given that the low number of iterations in the \emph{MCMC} chain prevents a high quality assessment of confidence interval, only a rough $\chi^{2}$ colourmaps in the remaining panels of Figure~\ref{fig:5} -- with the pink hexagon displaying the best fit location -- can be derived. These colourmaps suggest that good models present a slight degeneracy between decreasing planet eccentricity with increasing semi-major axis, as well as illustrates that good models are clustered around ($a_{p}=31~$AU, $e_{p}=0.2$) but can have a broad range of planetary masses with $6.5<m_{p}<8$. However, we can determine the probability distributions standard deviation to estimate the best fit parameters and their uncertainties -- see Table \ref{Table4}.
\begin{table}
\renewcommand{\arraystretch}{1.0}
\caption{Best fit values and estimation of the confidence intervals.}
\label{Table4}
\centering
\begin{tabular}{ccc}
\hline
Parameters & Best fit &  68\% conf. inter. \\
\hline
$m_{p}$ ($M_{\rm J}$) & 7.76 & $\pm 0.7$\\
$a_{p}$ (AU) & 31.07 & $\pm 1.35$\\
$e_{p}$ & 0.20 & $\pm 0.05$\\
\hline
\end{tabular}
\end{table} 
While increasing the number of iterations would help derive stronger confidence intervals, we decided to keep the number of iterations low for several reasons. Firstly, running dynamical simulations are computationally expensive (a run of 720 simulations requires $\sim 14$ days with a eight-cores Intel Xeon E5-2660 CPU of 2.2~GHz) and as a consequence, very few dynamical studies of debris discs in the literature use a set of more than a couple of hundred simulations~\citep{2014MNRAS.438.3577T,2014MNRAS.443.2541P,2014A&A...563A..72F}. To further ensure that enough planetary models resulting in disc parameters comparable to \cite{2015ApJ...807L...7C} were explored within our limited number of iterations, we compare the disc parameters achieved by planetary models along the \emph{MCMC} chain 1 with the observational parameters derived by \cite{2015ApJ...807L...7C} and their associated uncertainties in Figure~\ref{fig:C1}~\footnote{We use the authors' uncertainty on $r_{0}=\pm 1.1~$AU to derive the uncertainty on the disc width, which led to an uncertainty of 0.05 on the observational $\Delta r/r_{0}$}. After a hundred iterations, most planetary models resulted in disc parameters similar to the observational value within the observational uncertainty, and we therefore conclude that given the uncertainty associated with the observational parameters, our number of iterations is enough to probe the best-fit region with appropriate accuracy. Finally, direct imaging is currently the only technique capable of detecting planets with semi-major axes $>30~$AU. Checking the exoplanet database\footnote{http://exoplanetarchive.ipac.caltech.edu, 22/03/2016} for the 8 confirmed exoplanets which have been detected via imaging and with semi-major axes $20<a_{p}<60~$AU, we found that, when available, the average of uncertainties on the planetary parameters are $a_{p}=\pm 3.5$~AU and $m_{p}=^{+5.6}_{-1.9}~M_{\rm J}$. Our short \emph{MCMC} runs are therefore sufficient to also provide better constraints than those derived from current observations.

\begin{figure*}
\begin{center}
 \subfloat{\includegraphics[width=90mm,height=67mm]{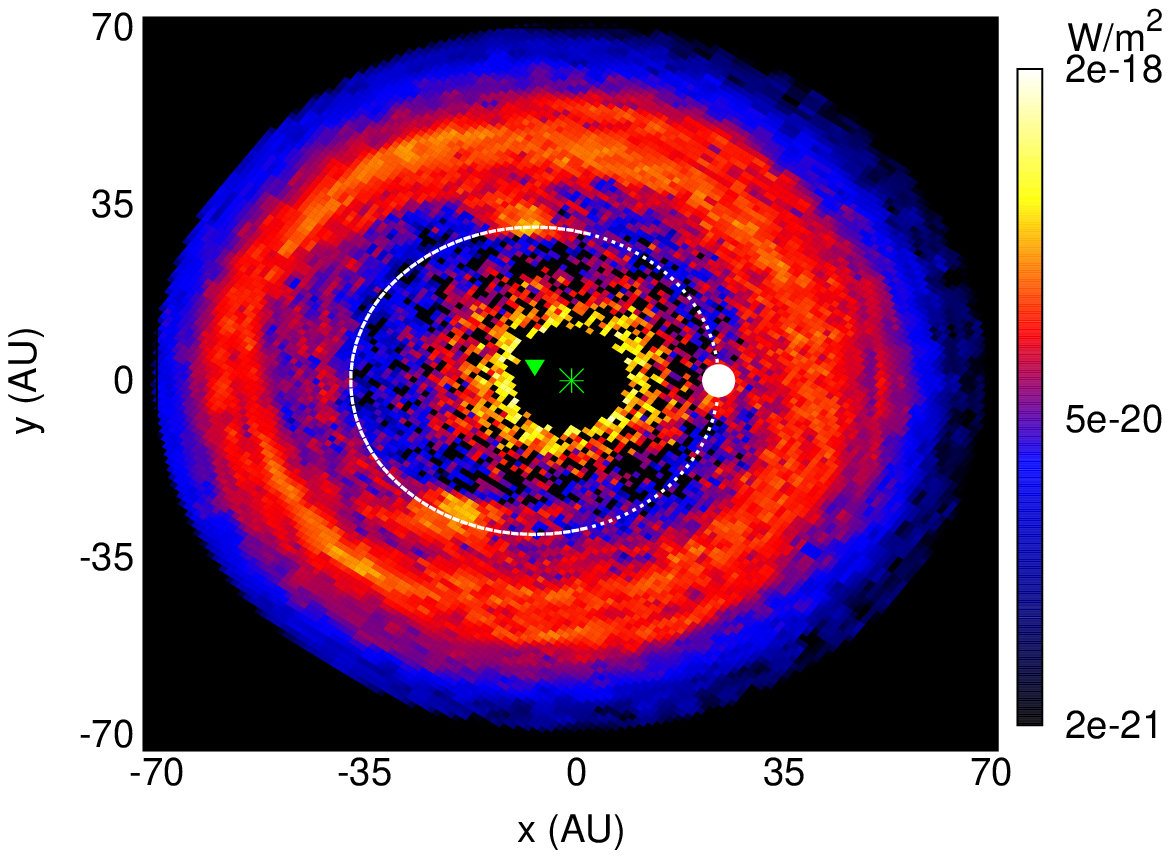}}
 \subfloat{\includegraphics[width=90mm,height=67mm]{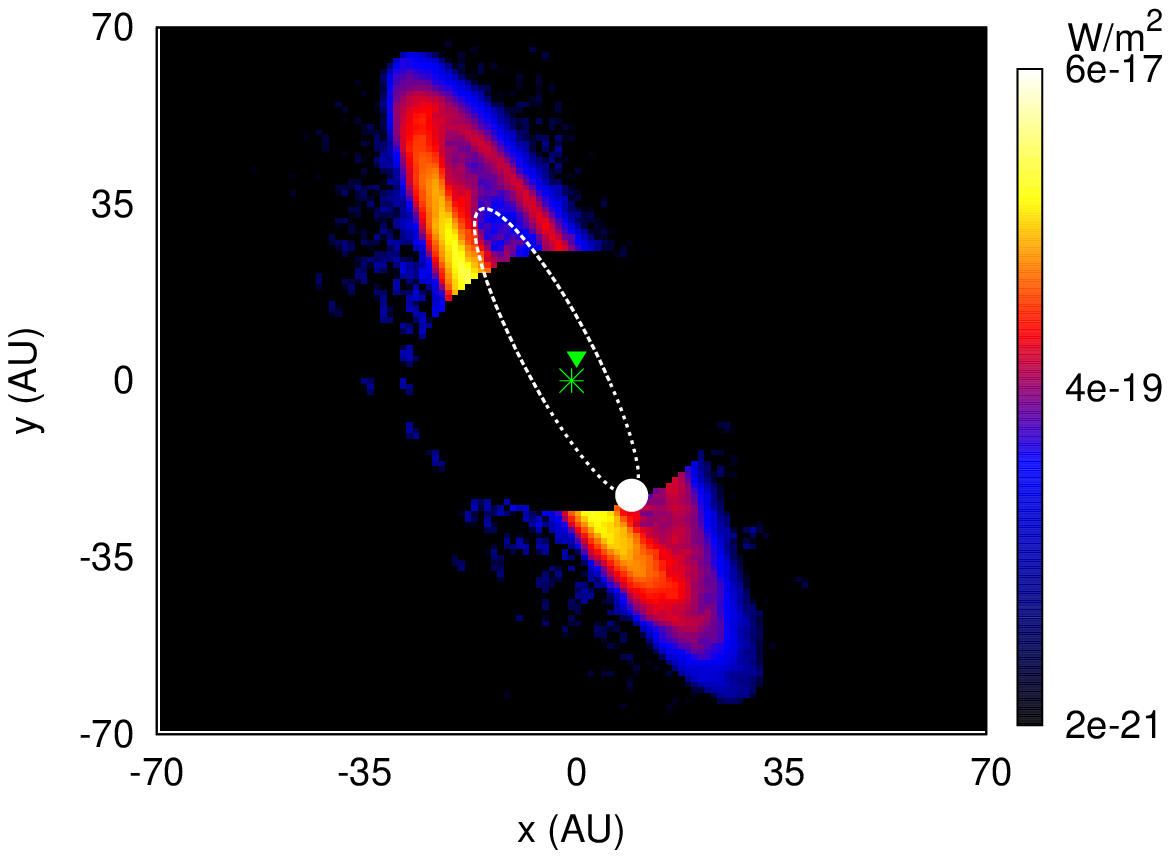}}  \\
 \subfloat{\includegraphics[width=90mm,height=67mm]{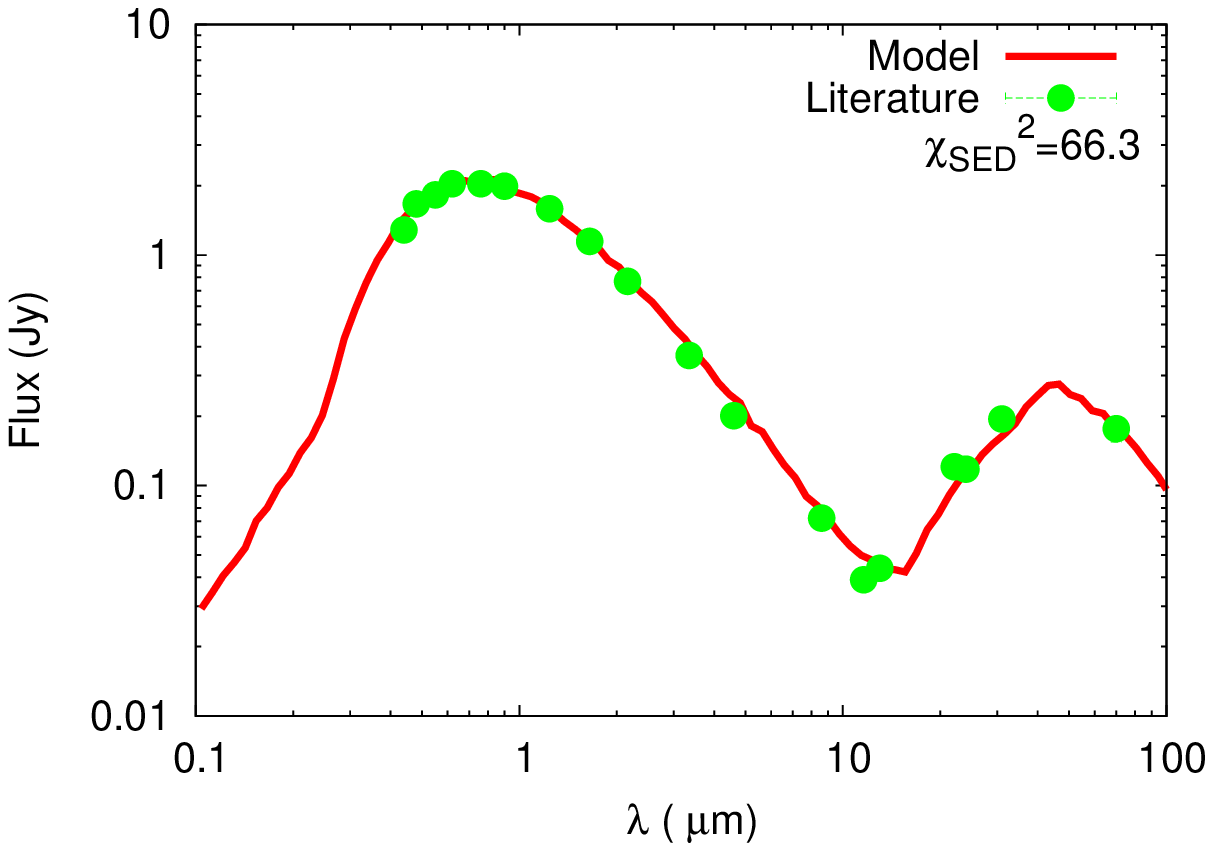}}                   
  \caption{(Left) \emph{MCFOST} deprojected synthetic image at 1.6~$\mu$m of the best fit model. The green star indicates the location of the star, while the disc offset centre is represented by the green triangle. The planet's orbit is represented by the white line and the planet's pericentre is indicated by the white dot. (Right) \emph{MCFOST} projected synthetic image at 1.6~$\mu$m of the best fit model using a mask of radius 26 AU, equivalent to the coronograph mask used in the \emph{GPI} image of Currie et al. (2015). (Bottom) \emph{MCFOST} spectral energy distribution of our best fit model compared with data from the literature (Table~\ref{Table2}).} 
\label{fig:6}  
\end{center}
\end{figure*} 

Using \emph{MCFOST}, we create the spectral energy distribution and synthetic image at 1.6~$\mu$m corresponding to the best fit model -- see Figure \ref{fig:6}. The SED of the best fit model has $\chi_{SED}^{2}=66.3$, which is larger than the $\chi^{2}$ of the best fit model using a parametric disc in Section 3 to estimate the grain properties. The high $\chi_{SED}^{2}$ value comes from the fact that we are trying to reproduce the disc features and emission observed in the near-IR and therefore our model, which is using small grains as they are the main contributors to the emission at short wavelengths, fails to reproduce the disc emission at the mid-IR wavelength (10-25 $\mu$m) accurately. On the other hand, the model reproduces the observed SED for $\lambda < 10~\mu$m very well. The projected synthetic image at $\lambda=1.6~\mu$m produced a disc with a projected offset of $\delta=4.95$~AU, and the deprojected disc has a peak brightness location of $r_{0}=47.8~$AU, a disc width ratio of $\Delta r/r_{0}=0.37$, and an eccentricity of $e=0.12$, which is a good match to the observed characteristics of Table \ref{Table1} within the uncertainties.

\begin{figure*}
\begin{center}
  \subfloat[][Grid method]{\includegraphics[width=90mm,height=67mm]{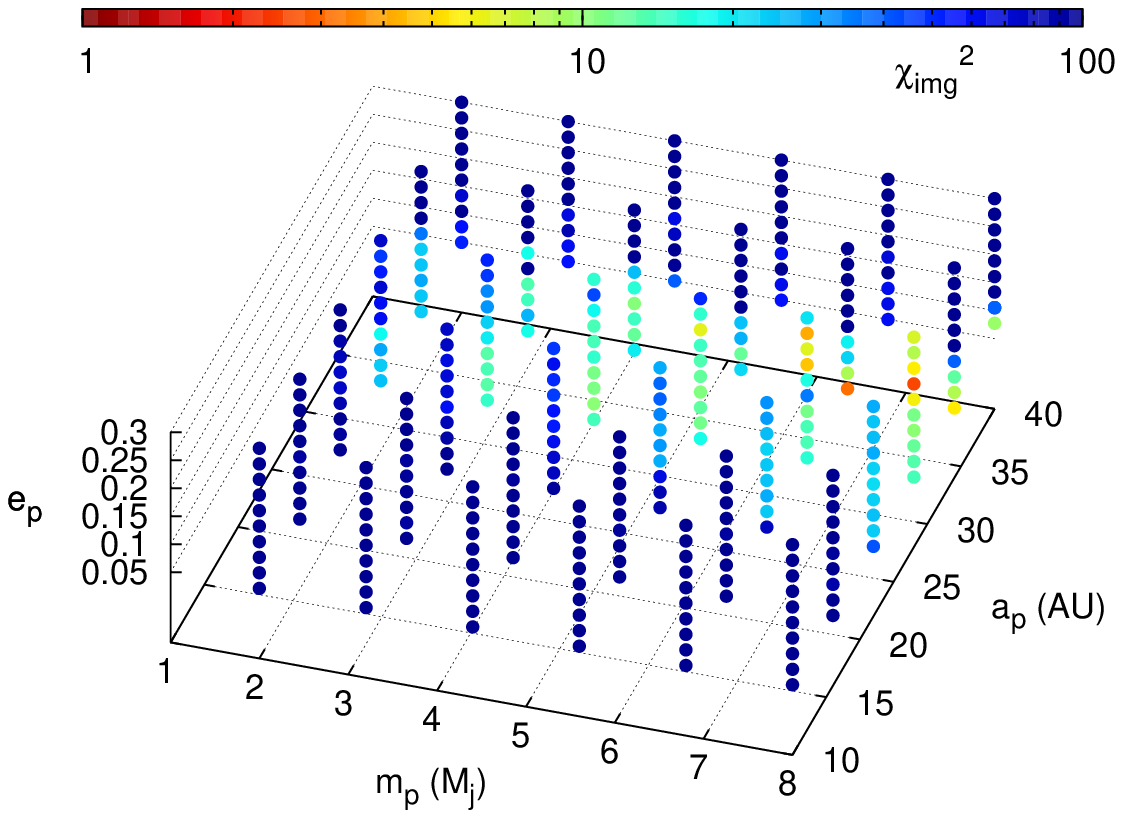}}  
   \subfloat[][MCMC method]{\includegraphics[width=90mm,height=67mm]{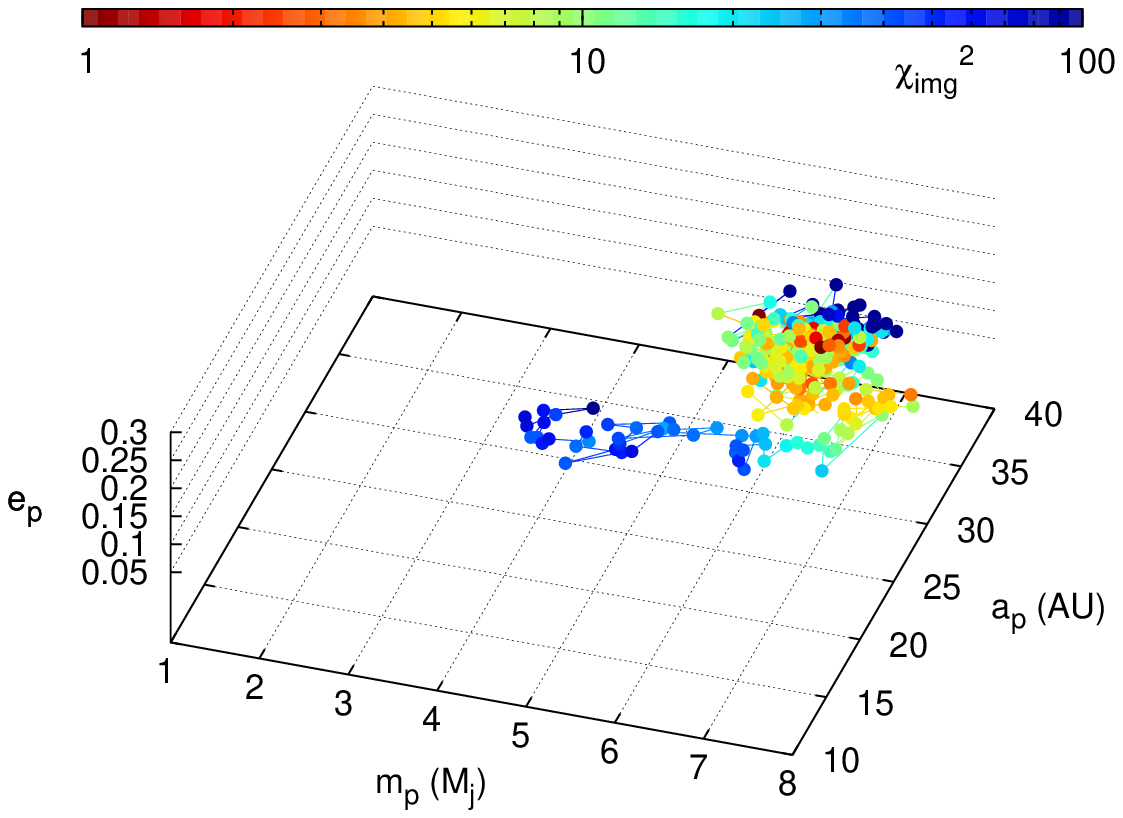}}                               
  \caption{$\chi^{2}$ map of the 3D parameter space explored in the 360 simulations using (a) the grid sampling method and (b) \emph{MCMC} sampling method. The colourbar indicates the $\chi_{img}^{2}$ value for each model.} 
\label{fig:7}  
\end{center}
\end{figure*}

\subsection{MCMC vs grid sampling}
To further check our implementation of the \emph{MCMC} sampling method, we conduct an exploration of the parameter space using a traditional grid sampling method. We sample the parameter space as follows: we used 6 values of $a_{p}$ between 10--40 AU in steps of 5~AU; 10 values of $e_{p}$ between 0.05 and 0.35 in steps of 0.06; and 6 values of $m_{p}$ between 2 $M_{\rm J}$  and 8 $M_{\rm J}$ in steps of 1.2. Those ranges corresponds to the values typically explored in the \emph{MCMC} chains (see Table~\ref{Table3}). A total of 360 models were conducted, and we aim here to compare the best fit model and the statistics of the result found when using a grid sampling method versus a \emph{MCMC} method for which only the first 360 iterations minus the 100 for the burn-in period are used.

\begin{figure*}
\begin{center}
  \includegraphics[width=188mm,height=130mm,trim=0mm 50mm 0mm 0mm,keepaspectratio=true,clip=true]{Fig8.eps}\\
  \caption{Bayesian probability distribution for the parameters $a_{p}$ (left), $e_{p}$ (centre) and $m_{p}$ (right) for the set of 360 simulations using a grid sampling of the parameter space (dashed blue) and the 360 first iterations of the MCMC chain 1 (red).} 
\label{fig:8}  
\end{center}
\end{figure*} 

Following a grid method, the best fit ($\chi^{2}_{img}=2.55$) is found for the model with $a_{p}=28~$AU, $e_{p}=0.21$ and $m_{p}=8~M_{\rm J}$, while by using only the first 360 simulations of the \emph{MCMC} chain 1, a better fit ($\chi_{img}^{2}=0.70$) is found for $a_{p}=30.7~$AU, $e_{p}=0.19$ and $m_{p}=6.75~M_{\rm J}$. While the best fit values are in excellent agreement between the two methods, Figure \ref{fig:7} shows a $\chi^{2}$ map of the grid and illustrates how the grid method can be wasteful in terms of computational effort. More than 250 out of 360 simulations returned $\chi_{img}^{2}>60$, highlighting that $70\%$ of the time was spent outside of the best fit region, while the \emph{MCMC} chain converged quickly toward the best fit region with all simulations having $\chi^{2}=7.56$ after the first 100 iterations. 

As a result, only poor statistics can be derived by following a grid method: Figure \ref{fig:8} presents the marginalized probability distribution of each parameter. While the planet mass, eccentricity and semi-major axis distribution peak corresponds to the best fit model, the $\sigma$ interval confidence for $m_{p}$ is $> 1.2 M_{\rm J}$ and no interval can be derived for $a_{p}$ or $e_{p}$. If we look now at the probability distribution of each parameter derived using the first 260 iterations of the \emph{MCMC} chain 1 (the first 360 simulations excluding 100 iterations as a burn-in period) presented as the red distribution in Figure \ref{fig:8}, the distribution for all parameters peaks around the best fit value with $\sigma_{m_{p}}\sim 0.3 M_{J}$, while no confidence interval can be derived for $a_{p}$ and $e_{p}$. Although these \emph{MCMC} probability distributions clearly provide less constraints on the interval confidence than the distribution derived using the full MCMC iterations sample (Figure \ref{fig:5}), it still presents slightly more constraints on the planetary mass confidence interval than the interval derived using the grid method. 

\subsection{Dynamical shaping of the disc}
\begin{figure}
\begin{center}
  \includegraphics[width=90mm,height=70mm,trim=15mm 0mm 9mm 0mm,keepaspectratio=false,clip=true]{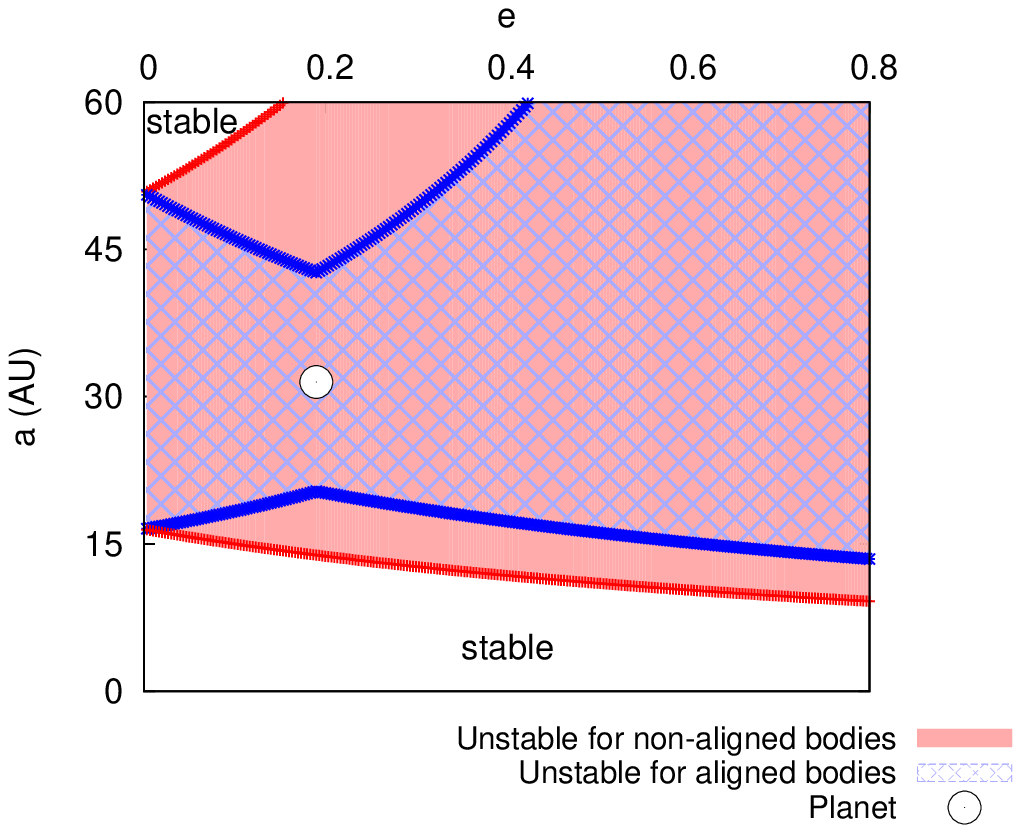}                            
  \caption{Theoretical stability map for the best fit model. The white dot represents the 7.6 Jupiter mass planet with $a_{p}=31$~AU and $e_{p}=0.19$, while the lines delimit the extended orbit crossing regions between the massive planet and a potential companion the mass of Vesta when the asteroid is either apse aligned with the planet (red) or not (blue dashed).} 
\label{fig:9}  
\end{center}
\end{figure}

We now examine the dynamics at work that is shaping the disc. The shaping of the debris disc by the planet is done via a mix of secular and resonant interactions, where secular interactions shape the disc eccentricity and orientation, while resonant interactions create dust traps and dictate their location in the disc. To understand and analyze the stability state of this system, we first use the criteria from \cite{2013MNRAS.436.3547G} to estimate the width of the chaotic zone around a planet with the best fit parameters for an asteroid of the mass of the moon used as a proxy for the total mass of the disc -- see Figure \ref{fig:9}.  This criteria predicts than in the absence of planet-disc apse alignment, the outer stable zone  around the planet is restricted to a small portion of the plot with $a>45~$AU and $e<0.2$, while if the planet and disc are aligned, the stable zone extend to $a > 35~$AU with $e<0.5$.

\begin{figure}
\begin{center}
  \includegraphics[width=0.35\textwidth,trim=0mm 0cm 88mm 0mm,keepaspectratio=true,clip=true]{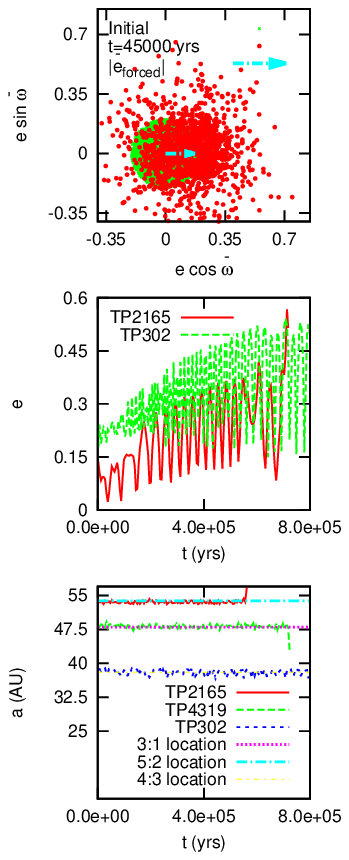}                            
  \caption{(Top) Complex eccentricity map ($e\cos{\overline{\omega}}$, $e\sin{\overline{\omega}}$) of all test particles constituting the disc at $t=0$ (green) and at $t=45000~$yrs (red). The blue arrow shows the location and direction of the forced eccentricity, $e_{forced}$, of the particles population at $t=45000$ yrs. (Bottom) Evolution plot of the semi-major axis and eccentricity of three test particles trapped in a 4:3, 3:1 and 5:2 MMR with the planet. } 
\label{fig:10}  
\end{center}
\end{figure}

At the beginning of the simulations, since the disc and planet are not initially aligned, particles with $a<45$ AU and $e>0.2$ are rapidly removed as they lie within the chaotic zone highlighted by the red region of Figure \ref{fig:9}. However, after $t \sim 25,000$ years (which corresponds to $0.7~t_{sec}$ for this $7.76~M_{\rm J}$ planet at 31.07 AU to sculpt the disc at 48 AU), secular interactions start to align the disc and planet. As a consequence, the chaotic zone shrinks to the blue region of Figure \ref{fig:9}, allowing the particles to evolve over a broader range of semi-major axes and eccentricities. 30\% of the initial test particles population has survived by this stage. The top panel of Figure \ref{fig:10} illustrates the impact of secular interaction on the disc eccentricity and orientation. Initially the particles' complex eccentricities ($e\cos{\overline{\omega}}$, $e\sin{\overline{\omega}}$) are restricted to the green disc of radius $e=0.2$ centred on the origin, which corresponds to particles having an initial eccentricity range between $0<e<0.2$ (as set in the initial conditions). By $t>0.7~t_{sec}$, the complex eccentricities are now precessing about the forced eccentricity, $e_{forced}=0.17$, as indicated by the blue arrow and form the red disc of radius $e_{free}=e_{forced}$. In addition, because the planet is initially set with its pericentre $\overline{\omega_{p}}=0$, the direction of the disc forced pericentre illustrated by the blue arrow is the $x$-axis, i.e., $\overline{\omega_{forced}}=0$. The disc and planet are therefore aligned.

In addition to secular interactions, the disc location is also shaped by a set of resonances due to the interaction with the planet, which result in the presence of dusty clumps in the deprojected synthetic image of the best fit disc model in Figure~\ref{fig:6}. At the beginning of the simulation, while 70\% of the test particles are scattered by the planet, the surviving particles are clustered around 38.5, 48 and 54 AU, which correspond to the 4:3, 3:1 and 5:2 mean motion resonances. The evolution of the semi-major axis and eccentricity of two particles trapped in the 4:3, 3:1 and 5:2 mean motion resonances is shown in Figure \ref{fig:10}. The 3:1 MMR is trapping test particles at $a=48~$AU and, as a result, the particles' eccentricity is undergoing adiabatic growth, while the 5:2 MMR is trapping particles around 54 AU. These resonances induce stability in the disc and the final disc is made of long term surviving particles with semi major axis mainly clustered around these MMRs with an eccentricity $0<e<0.3$.
 
The new generation of high contrast imaging instruments, such as the \emph{GPI}, can reach contrasts of $10^{-5}$ within the inner $0.4^{''}$. Using the isochrones computed by \cite{2003A&A...402..701B} for the evolutionary models of brown dwarfs and giant exoplanets applied to a 7 Jupiter mass of a same age than HD~115600, we estimate the flux ratio contrast to be $4\times10^{-5}$ in the H band. While technically within the current detection threshold, there are several reasons why our best fit planet was not detected on the Currie et al. image. First, given the high inclination of the system ($i=79^{\circ}$), the apparent orbit of the planet may be hidden by the disc emission, and secondly, most of the planet orbit is likely concealed by the coronograph mask which has an apparent radius of $26~$AU, as can be seen on the right panel of Figure \ref{fig:6}. We however note that a small portion of the planet's orbit around the apocentre is situated beyond the mask, and the planet at this position could be marginally detectable.

\section{Discussion \& Conclusions}
In this work, we conducted a numerical search for a planet responsible for shaping the debris disc of HD~115600 recently imaged in the H band by \cite{2015ApJ...807L...7C} using \emph{GPI}.

We first use the radiative transfer code \emph{MCFOST} with a parametric disc structure to derive the dust grain properties needed to fit the observed SED in order to create synthetic images. We then used a modified N-body integrator which incorporates radiation forces to dynamically model the interaction between the potential companion and the debris disc, and used \emph{MCFOST} to produce synthetic images to compare with observations. We explored the 3D parameter space ($a_{p}$,$e_{p}$ and $m_{p}$) of the potential planet using two different methods: a classic grid exploration over 360 grid points, and a \emph{MCMC} approach over 720 iterations. Our main results are as follows:
\begin{itemize}
\item From our SED modeling, and by selecting the Greenburg model where the grains are made of a core of silicate with a coating of water ice, the disc emission is best reproduced by a disc with a total dust mass of 0.2~$M_{\rm moon}$ with a minimal and maximal grain size of $0.05~\mu$m and $75~\mu$m. We however note that these parameters are not well constrained.
\item From image modeling and using the \emph{MCMC} scheme to explore the parameter space, we reproduce the HD115600 disc eccentricity, projected offset, disc width and brightness peak with a $7.76~M_{\rm J}$ planet located at $a_{p}=31.07~$AU with $e_{p}=0.20$. Such a planet shapes the disc via a mix of secular interactions forcing the disc eccentricity, orientation and offset, as well as a set of planet resonances, at 4:3, 3:1 and 5:2, trapping dust at $a=38.5$, $48$ and $54$ AU respectively.
\item While technically detectable by \emph{GPI}, the apparent and projected planet orbit is likely hidden by the disc emission as well as partially concealed by the coronograph mask due to the high inclination of the system.
\item Using a same total number of simulations, we compared two methods of exploring the parameter space: a \emph{MCMC} sampling of the parameter space versus a grid method. We find that the \emph{MCMC} scheme not only finds a better fit to the observations, but also converges faster toward the best fit region.  However we find that using a \emph{MCMC} scheme on such a short number of iterations presents some limitations: while our \emph{MCMC} implementation barely passed the convergence test such as the \cite{1992GelmanRubin}, it also only provides weak constraints on the confidence intervals. Increasing the numbers of iterations would increase the quality of the best fit estimation. We chose to conduct our study with a small number (720) of simulations for multiple reasons: first very few studies in the literature uses more than a few hundred dynamical simulations as they are computationally expensive and, secondly, after about a hundred iterations, the majority of subsequent models resulted in disc parameters comparable to the values obtained by \cite{2015ApJ...807L...7C} within the observational uncertainties. Finally our planetary parameters uncertainties estimated with 720 \emph{MCMC} iterations are well below the current uncertainties of parameters for planets detected via direct imaging.
\end{itemize}
For low dimensional problems (low $D<2-3$), like our study, a traditional grid of $M$ points only requires $M^{D}$ evaluations and presents the advantage of being trivially parallelized. For higher dimensional problems, it should be noted the \emph{MCMC} efficiently converges like $\sqrt{1/M}$, which is dimensionally independant and therefore a large advantage over the grid technique. In this study, we compare the efficiency of exploring our $D=3$ parameter space with a grid and a \emph{MCMC} algorithm. We conclude that this work demonstrates that the \emph{MCMC} scheme is a promising method to efficiently explore the parameter space of dynamical simulations and allows us to localize a better-fit region faster than the grid method, although we also stress that the limited number of simulations makes reaching the convergence state quite challenging. There are numerous schemes existing in the literature to explore multidimensional parameter spaces which could represent very good alternatives, such as the \emph{sparse grid MCMC} \citep{Menze2011} -- a scheme combining the use of both a grid and \emph{MCMC}, or the use of grids of different resolutions successively \citep{2013MNRAS.435.2033H}. The debris disc of HD~115600 shows interesting features which can be reproduced by simulating the gravitational influence of an inner massive planet on the disc structure. Additional observations of HD~115600 at different wavelengths could help constraining the planetary parameters furthermore.

\section*{Acknowledgements}
This work was performed on the swinSTAR supercomputer at Swinburne University of Technology. E.T. would like to thank Chris Blake and Caitlin Adams for useful discussions about the \emph{MCMC} algorithm, as well as Thayne Currie and Tadashi Mukai for providing information about the grain composition and modeling. We also thank the anonymous referee for constructive feedback which have helped improve this paper. E.T. was supported by a Swinburne University Postgraduate Research Award (SUPRA).


\appendix

\section{The recording process}
The small dust grains experience orbital perturbation by stellar radiation forces and are short-lived. Collisions are the assumed mechanism by which new dust is constinuously created, and following our discussion in Section 4.1.2, we assume the debris disc of HD~115600 to be either (i) in a steady-state where larger planetesimals collide and constinuously replenish the disc with small grains or, given the young age of the system, (ii) is replenished by collisions of primordial grains before all are removed by stellar radiation forces. Either way, collisions are expected to occur within HD~115600 debris disc on a timescale shorter than the simulation duration, and by stacking the dust distribution over the simulation duration, at any time $t$ we therefore mimic the dust distribution for grains whose ages range from $t$ years old to those newly created.

\begin{figure*}
\begin{center}
 \subfloat{\includegraphics[width=60mm,height=45mm]{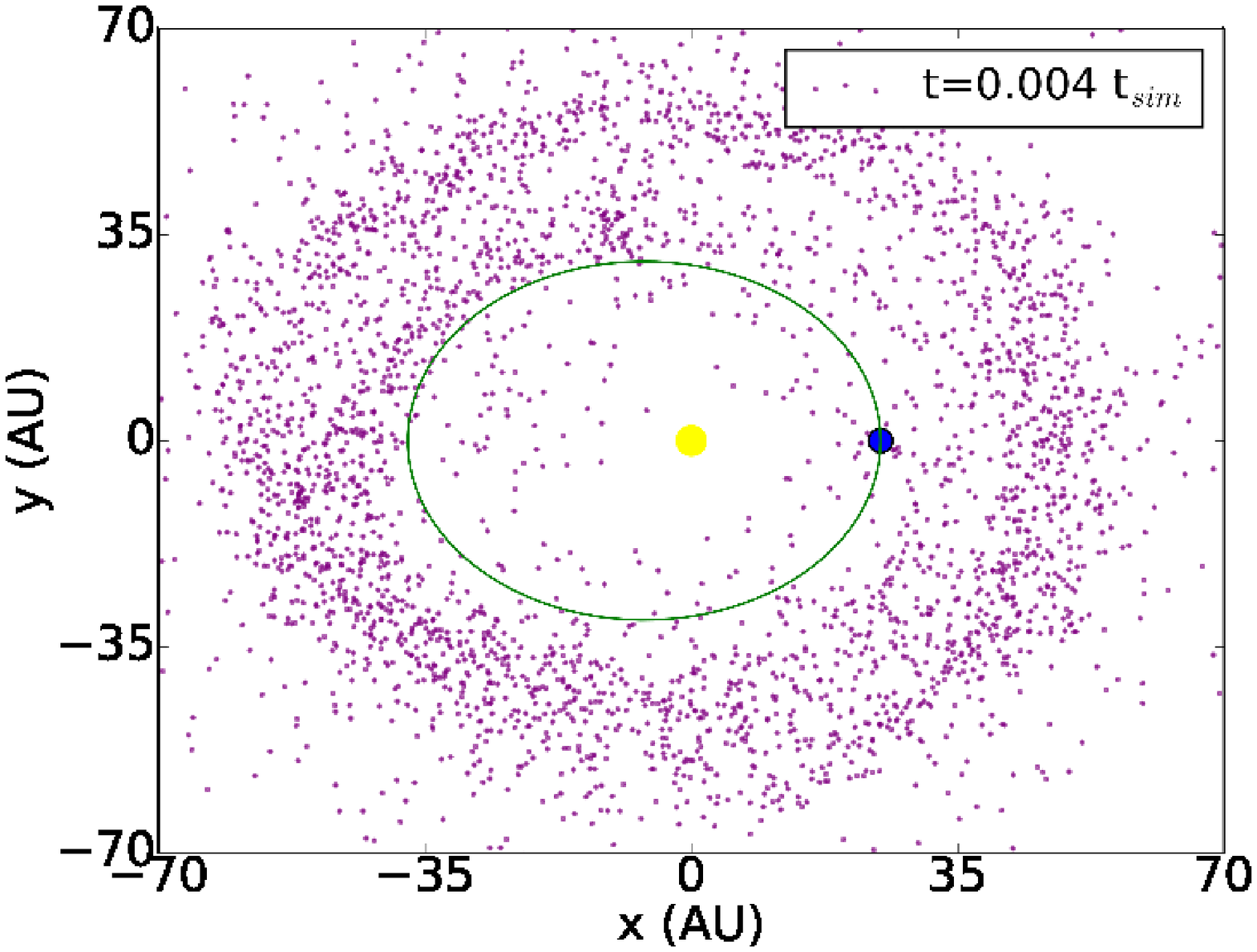}}  
 \subfloat{\includegraphics[width=55mm,height=45mm]{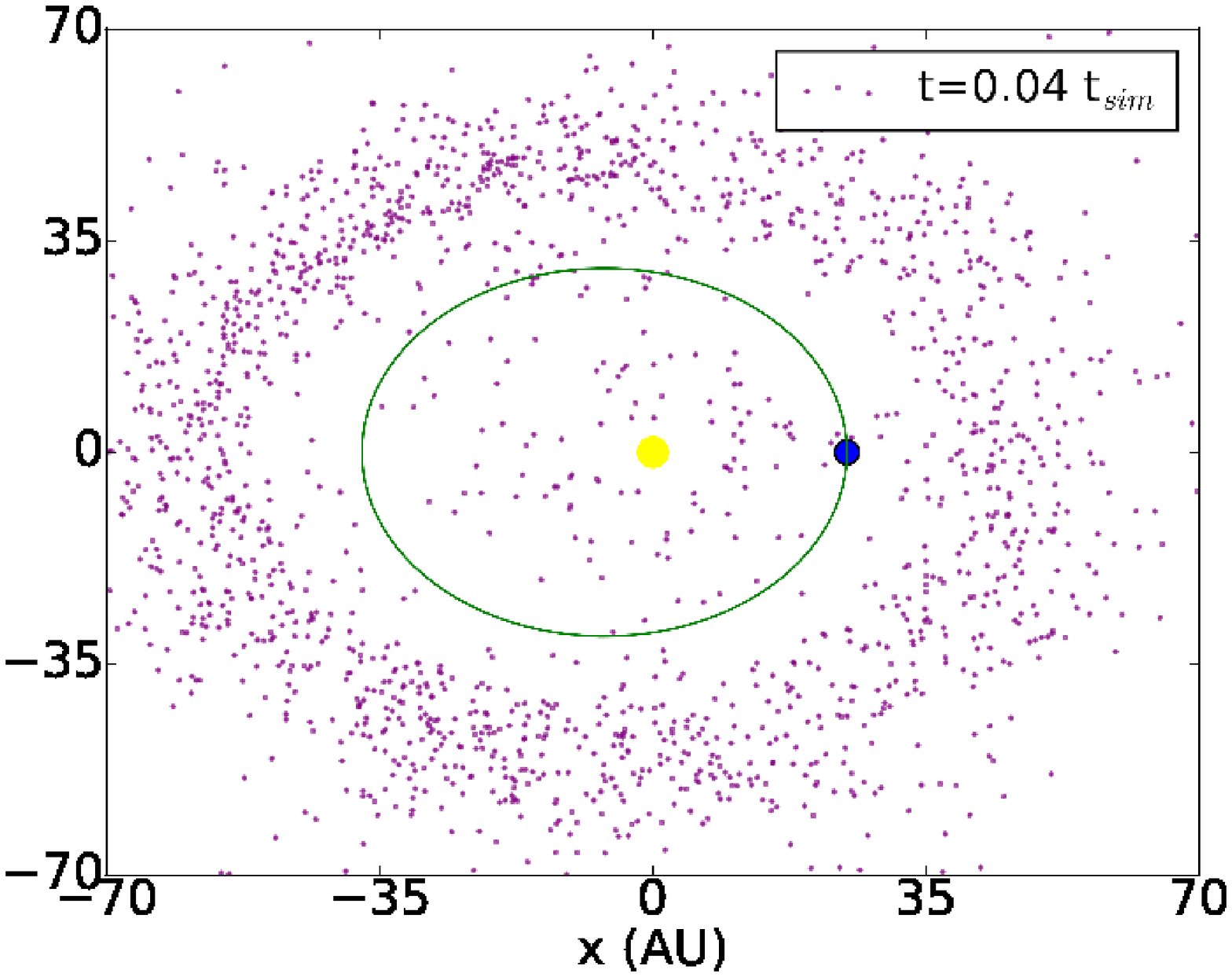}}      
 \subfloat{\includegraphics[width=55mm,height=45mm]{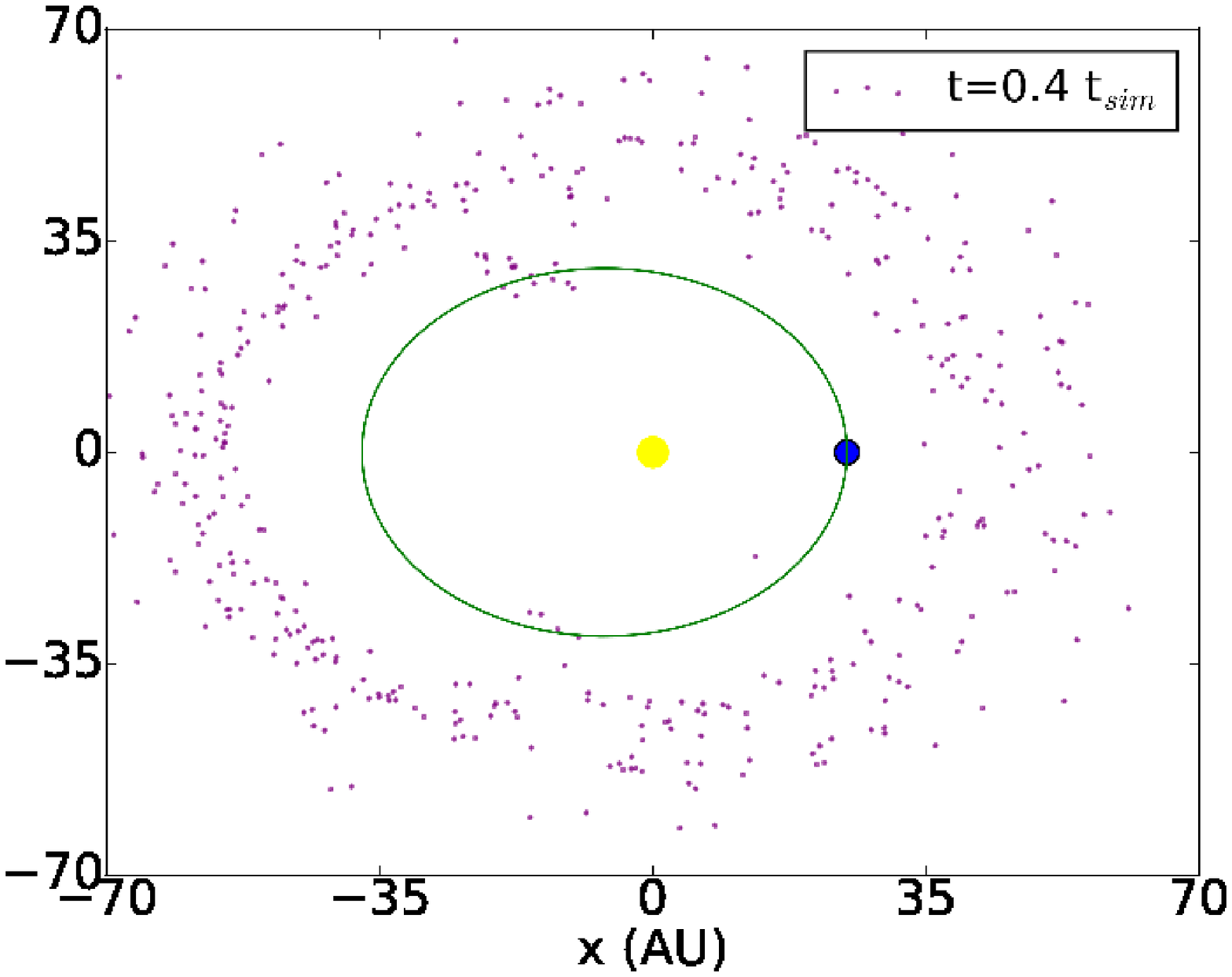}}                      
 \caption{Three evolutionnary snapshots of the dust distribution of the best fit model simulation of the \emph{MCMC} chain 1. The yellow and blue dots correspond to the star and planet location respectively, while the green line represents the planet orbit.} 
\label{fig:A1}  
\end{center}
\end{figure*}

It is valid to then ask if mimicing the effect of collisions smoothens out the planet-induced features in the disc and therefore impacts the resulting synthetic disc image (as well as the subsequent determination of the disc parameters). In Figure~\ref{fig:A1}, snapshots of the dust distribution at the early stage of a simulation ($t=0.004~t_{sim}$) shows a broad slighly perturbed disc, while later snapshots taken over $t>0.04~t_{sim}$ (or 45000~years) all display a clear eccentric ring that is apse-aligned with the planet. By $t=0.4~t_{sim}$, the presence of two dust clumps corresponding to MMR trapping can be clearly distinguished and little disc evolution occurs after that. Interestingly, the structure observed at $t=0.4~t_{sim}$ is consistent with the disc structure illustrated in the deprojected \emph{MCFOST} synthetic image in Figure~\ref{fig:6}, for which the dust density distribution was obtained by stacking all the snapshots over the lifetime of the simulations. Therefore, because both the secular interactions and resonnance trapping occur rapidly within the disc (see also Figure~\ref{fig:10}), no significant difference is expected between the dust density distribution at the end of the simulations and the dust distribution obtained by stacking.

\section{Tuning the \emph{MCMC}}
A quick way to verify that the parameter space has been efficiently sampled is to have an overall acceptance rate of 25 to 40\% in the \emph{MCMC} chain. This requires adjusting the standard deviation of the three normal prior parameter distributions, from which a new set of parameters is drawn at every \emph{MCMC} step. A commonly used method is to have a some correlation stepping when drawing a new set of parameters at each \emph{MCMC} iteration, meaning that the value of the standard deviation for each prior distribution is based on how the three parameters are correlated. In this section, we provide a short description on how to achieve this but we refer to \cite{2005MNRAS.356..925D} for more details.

\begin{enumerate}
\item Step 1: The method consists in initially setting the normal prior distributions with a standard deviation of $\sigma_{a_{p}1}=\sigma_{e_{p}1}=\sigma_{m_{p}1}=1$ for all parameters, and to run the \emph{MCMC} chain for a few hundreds iterations. 
\item Step 2: By recording how the chain moves through the parameter space, we can estimate the covariance matrix, \textbf{$C_{1}$}, of the three parameters and therefore know how the sets of parameters over all \emph{MCMC} iterations are correlated.
\item Step 3: The next stage is to adjust the standard deviation of the prior distributions so that the sets of parameters would follow the covariance we just estimated, and this requires to generate correlated random numbers. This process can be easily done by first performing a Cholesky decomposition on the covariance matrix, which consists in finding an upper diagonal matrix \textbf{$U_{1}$} such that \textbf{$C_{1}$}=\textbf{$U_{1}^{T}U_{1}$}. 
\item Step 4: The Cholesky matrix, \textbf{$U_{1}$}, then needs to be updated to \textbf{$U_{U1}$} to obtain a better acceptance rate, where \textbf{$U_{U1}$}$=(2.4^{2}/D)\times$\textbf{$U_{1}$}, with $D=3$, the number of parameters constituting the parameter space. 
\item Step 5: The new standard deviation for each parameters, $\sigma_{a_{p}2}, \sigma_{e_{p}2}$ and $\sigma_{m_{p}2}$ is then obtained by applying \textbf{$U_{U1}$} to the vector ($\sigma_{a_{p}1},\sigma_{e_{p}1},\sigma_{m_{p}1}$).
\item Step 6: A new \emph{MCMC} chain can be run using the correlated stepping $\sigma_{a_{p}2}, \sigma_{e_{p}2}$ and $\sigma_{m_{p}2}$, to generate the set of parameters at each iteration of the chain.
\end{enumerate}
If the new acceptance rate is not inside the ideal range, then steps 2 to 6 must be repeated until the acceptance rate enters the correct range. 

\section{Traceplot of the disc parameters}
\begin{figure*}
\begin{center}
  \includegraphics[width=120mm,height=107mm]{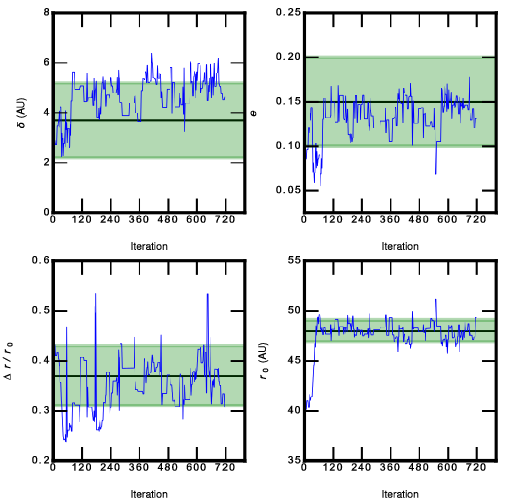}                            
  \caption{Traceplot of the 4 disc parameters ($\delta$, $e$, $\Delta r/r_{0}$ and $r_{0}$) within the \emph{MCMC} chain 1. The black line represents the observational value for each parameter as determined by Currie et al. (2015), and the green region represents the uncertainties associated with each observational value.} 
\label{fig:C1}  
\end{center}
\end{figure*} 

\label{lastpage}
\end{document}